\documentclass[a4paper,11pt]{article}
\pdfoutput=1 

\usepackage{jcappub} 

\usepackage[T1]{fontenc} 

\usepackage{aasmacros}

\usepackage{amsmath}
\usepackage{bm}
\usepackage{tabularx}
\usepackage{mleftright}
\usepackage{xfrac} 
\usepackage{booktabs}

\newcommand{\Cos}[1]{ \mathrm{c}({#1}) }
\newcommand{\Sin}[1]{ \mathrm{s}({#1}) }
\newcommand{\Tan}[1]{ \mathrm{t}({#1}) }
\newcommand{\CosS}[1]{ \mathrm{c}^2({#1}) }
\newcommand{\SinS}[1]{ \mathrm{s}^2({#1}) }

\usepackage{soul} 

\usepackage{caption}
\usepackage{subcaption}

\usepackage{xcolor}
\definecolor{columbiablue}{rgb}{0.61, 0.87, 1.0}
\usepackage[colorinlistoftodos,color=columbiablue]{todonotes}

\makeatletter
\newcommand\notsotiny{\@setfontsize\notsotiny\@vipt\@viipt}
\makeatother

\title{Robustness of cosmic birefringence measurement against Galactic foreground emission and instrumental systematics}
\author[1,2]{P. Diego-Palazuelos}
\author[1]{E. Mart\'inez-Gonz\'alez,}
\author[1]{P. Vielva,}
\author[1]{R. B. Barreiro,}
\author[3]{M. Tristram,}
\author[1,2]{E. de la Hoz,}
\author[4]{J. R. Eskilt,}
\author[5]{Y. Minami,}
\author[6]{R. M. Sullivan,}
\author[7]{A. J. Banday,}
\author[8,9]{K. M. G\'orski,}
\author[10,11]{R. Keskitalo,}
\author[12,13]{E. Komatsu,}
\author[6]{and D. Scott}

\affiliation[1]{Instituto de F\'isica de Cantabria (CSIC - Universidad de Cantabria), Avenida los Castros s/n, 39005 Santander, Spain}
\affiliation[2]{Departamento de F\'isica Moderna, Universidad de Cantabria, Avenida de los Castros s/n, 39005 Santander, Spain}
\affiliation[3]{Universit\'e Paris-Saclay, CNRS/IN2P3, IJCLab, 91405 Orsay, France}
\affiliation[4]{Institute of Theoretical Astrophysics, University of Oslo, Blindern, N-0315 Oslo, Norway}
\affiliation[5]{Research Center for Nuclear Physics, Osaka University, Ibaraki, Osaka 567-0047, Japan}
\affiliation[6]{Department of Physics \& Astronomy, University of British Columbia, 6224 Agricultural Road, Vancouver, British Columbia, Canada}
\affiliation[7]{IRAP, Universit\'e de Toulouse, CNRS, CNES, UPS, Toulouse, France}
\affiliation[8]{Jet Propulsion Laboratory, California Institute of Technology, 4800 Oak Grove Drive, Pasadena, California, USA}
\affiliation[9]{Warsaw University Observatory, Aleje Ujazdowskie 4, 00-478 Warszawa, Poland}
\affiliation[10]{Computational Cosmology Center, Lawrence Berkeley National Laboratory, Berkeley, California 94720, USA}
\affiliation[11]{Space Sciences Laboratory, University of California, Berkeley, California 94720, USA}
\affiliation[12]{Max Planck Institute for Astrophysics, Karl-Schwarzschild-Strasse 1, D-85748 Garching, Germany}
\affiliation[13]{Kavli Institute for the Physics and Mathematics of the Universe (Kavli IPMU, WPI), Todai Institutes for Advanced Study, The University of Tokyo, Kashiwa 277-8583, Japan}

\emailAdd{diegop@ifca.unican.es}

\abstract{The polarization of the cosmic microwave background (CMB) can be used to search for parity-violating processes like that predicted by a Chern-Simons coupling to a light pseudoscalar field. Such an interaction rotates $E$ modes into $B$ modes in the observed CMB signal by an effect known as cosmic birefringence. Even though isotropic birefringence can be confused with the rotation produced by a miscalibration of the detectors' polarization angles the degeneracy between both effects is broken when Galactic foreground emission is used as a calibrator. In this work, we use realistic simulations of the High-Frequency Instrument of the \textit{Planck} mission to test the impact that Galactic foreground emission and instrumental systematics have on the recent birefringence measurements obtained through this technique. Our results demonstrate the robustness of the methodology against the miscalibration of polarization angles and other systematic effects, like intensity-to-polarization leakage, beam leakage, or cross-polarization effects. However, our estimator is sensitive to the $EB$ correlation of polarized foreground emission. Here we propose to correct the bias induced by dust $EB$ by modeling the foreground signal with templates produced in Bayesian component-separation analyses that fit parametric models to CMB data. Acknowledging the limitations of currently available dust templates like that of the \texttt{Commander} sky model, high-precision CMB data and a characterization of dust beyond the modified blackbody paradigm are needed to obtain a definitive measurement of cosmic birefringence in the future.  }

\begin{document}
\maketitle
\flushbottom

\section{Introduction}
\label{sec:intro}

Parity-violating processes are predicted by several extensions of the standard model of cosmology and particle physics~\cite{Gubitosi2013}. For example, axion-like particles~\cite{Marsh2016} and other dark energy and dark matter models~\cite{DM,DE} introduce a new parity-violating pseudoscalar field, $\phi$, that can couple to the electromagnetic tensor through a Chern-Simons interaction~\cite{Carroll1990,Carroll1991,Harari1992}. Such an interaction makes the phase velocities of the right- and left-handed helicity states of photons differ, rotating the plane of linear polarization clockwise on the sky by an angle $\beta = - \frac{1}{2}g_{\phi\gamma}\int \frac{\partial \phi}{\partial t} dt$ that depends on the coupling constant of the pseudoscalar field to photons $g_{\phi\gamma}$, and the time evolution of the field. This rotation is what we call ``cosmic birefringence'' because it is as if space itself acted like a birefringent material (see Ref.~\cite{Eiichiro_review} for a review). Cosmic birefringence can also be produced by the Faraday rotation originating from primordial magnetic fields~\cite{Campanelli2004,Subramanian2016}. Unlike the Chern-Simons interaction, Faraday rotation does depend on the photon energy, leading to a $\beta\propto\nu^{-2}$ birefringence angle. A frequency-dependent birefringence is also predicted by superluminal Lorentz-violating electrodynamics emerging from a non-vanishing Weyl tensor ($\beta\propto\nu$)~\cite{Shore2005}, and some quantum gravity models that modify the dispersion relation of photons ($\beta\propto\nu^{2}$)~\cite{Gleiser}. Nevertheless, in this work, we focus on the frequency-independent birefringence predicted by light pseudoscalar fields, since the analysis of \textit{Planck} data presented in Ref.~\cite{johannes} highly disfavored these other theories.

Although we know that birefringence must be a small effect, in principle we could constrain $\beta$ by measuring the rotation of the plane of polarization of a well-known source of linearly polarized light situated at a far enough distance to allow photons to accumulate a significant rotation. Emitted at the epoch of recombination and with its polarization angular power spectra accurately predicted by the $\Lambda$ cold dark matter ($\Lambda$CDM) model, the cosmic microwave background (CMB) is, therefore, the ideal tool in the search for cosmic birefringence~\cite{lue1999}.

We can model the effect of a constant, isotropic, and frequency-independent birefringence angle (like the one that a homogeneous axion-like field of mass $10^{-33}$eV$\leq m_\phi\leq 10^{-28}$eV might produce~\cite{fedderke2019, johannes, tomography_komatsu}) as a rotation of the plane of linear polarization of CMB photons. In this way, the spherical harmonic coefficients of the $E$ and $B$ modes of the CMB polarization that we observe (``o'' superscript) would be a rotation of those emitted at recombination:
\begin{equation}\label{eq:rotated_CMB}
    \begin{pmatrix}
		E_{\ell m}^{\mathrm{o}} \\
	    B_{\ell m}^{\mathrm{o}}
	\end{pmatrix}
	=
	\begin{pmatrix}
		\mathrm{c}(2\beta) & -\mathrm{s}(2\beta)\\
	    \mathrm{s}(2\beta) & \phantom{-}\mathrm{c}(2\beta)
	\end{pmatrix}	
	\begin{pmatrix}
		E_{\ell m}^{\mathrm{CMB}} \\
	    B_{\ell m}^{\mathrm{CMB}}
	\end{pmatrix}\, .
\end{equation}
For brevity, throughout this work we refer to the sine, cosine, and tangent functions as ``$\mathrm{s}$'', ``$\mathrm{c}$'', and ``$\mathrm{t}$'', respectively. Under this approximation, and without solving the Boltzmann equations coupled to the light pseudoscalar field (as done, e.g., in Ref.~\cite{tomography_komatsu}), we can model the observed angular power spectra as a rotation of the CMB spectra predicted in $\Lambda$CDM:
\begin{equation}\label{eq:cmb_EE_EB_BB_correlations}
    \begin{pmatrix}
        C_{\ell}^{EE, \mathrm{o}} \\
        C_{\ell}^{EB, \mathrm{o}} \\
        C_{\ell}^{BB, \mathrm{o}} 
    \end{pmatrix}
    = 
    \begin{pmatrix}
        \CosS{2\beta}  & -\Sin{4\beta} & \phantom{-}\SinS{2\beta} \\
        \frac{1}{2}\Sin{4\beta}  & \phantom{-}\Cos{4\beta} & -\frac{1}{2}\Sin{4\beta} \\
        \SinS{2\beta}  & \phantom{-}\Sin{4\beta} & \phantom{-}\CosS{2\beta} 
    \end{pmatrix}
    \begin{pmatrix}
        C_{\ell}^{EE, \Lambda\mathrm{CDM}}\\
        0\\
        C_{\ell}^{BB, \Lambda\mathrm{CDM}} 
    \end{pmatrix}\,.
\end{equation}
From Eq.~(\ref{eq:cmb_EE_EB_BB_correlations}) it follows that the observed $EB$ correlation can be written as a rotation of the observed $EE$ and $BB$ angular power spectra like
\begin{equation} \label{eq:observed_EB_CMB}
    C_\ell^{EB,\mathrm{o}} = \frac{\Tan{4\beta}}{2}\Big( C_\ell^{EE,\mathrm{o}} - C_\ell^{BB,\mathrm{o}}\Big).
\end{equation}

Eq.~(\ref{eq:observed_EB_CMB}) has been the basis for the majority of the harmonic-space methodologies applied in the past to measure cosmic birefringence from CMB polarization data~\cite{WMAP&BOOMERANG2006, QUaD2009, WMAP2013, Planck_CB_2016, ACTanisotropo2020, Polarbear2020, SPT2020, ACTisotropo2020}. However, those analyses have often been dominated by systematic uncertainties. In particular, the miscalibration of the detector’s polarization angle is one of the most pernicious systematics for this type of analysis, since it produces a rotation of the observed polarization signal that is degenerate with that of birefringence~\cite{Hu2003,Shimon2008,Miller2009,Yadav2010}. Namely, for an $\alpha$ miscalibration angle, the CMB spherical harmonic coefficients in Eq.~(\ref{eq:rotated_CMB}) would be rotated by $\beta + \alpha$, so that the observed $EB$ correlation in Eq.~(\ref{eq:observed_EB_CMB}) yields $\beta + \alpha$ instead of $\beta$. In this way, the calibration strategies used for currently available CMB datasets tend to limit the systematic uncertainty attainable through the analysis of $EB$ to $0.5^\circ$-$1^\circ$~\cite{Keating2013, Chiang2010, Naess2014, Planck_CB_2016, Sayre2020, Adachi2020}. In addition to the miscalibration of polarization angles, other systematic effects, like intensity-to-polarization leakage, beam leakage, or cross-polarization effects, also produce spurious $EB$ correlations that contribute to the total systematic uncertainty~\cite{Hu2003, Shimon2008, Miller2009, Yadav2010}. Although the accuracy in the calibration of polarization angles is expected to improve in the near future~\cite{PolAngle_requirements, calibration_Earth_satellite, calibration_Earth_balloon, calibration_Earth_ground, calibration_Space_satellite}, systematics will play an even more critical role in the precision measurements of CMB polarization envisioned for next-generation experiments~\cite{LiteBIRD_PTEP,simons_observatory,CMB_S4}.

To overcome the limitation imposed by the calibration of polarization angles, Refs.~\cite{Minami2019_auto, Minami2020cutsky, Minami2020cross} proposed a novel methodology to simultaneously determine birefringence and miscalibration angles through the use of polarized Galactic foreground emission. Foreground emission can be used to break the degeneracy between the $\alpha$ and $\beta$ angles since Galactic foreground photons are negligibly affected by cosmic birefringence due to their small propagation length. That methodology has proven to successfully capture polarization angle miscalibrations and provide robust birefringence measurements~\cite{Minami2019_auto, Minami2020cutsky,Minami2020cross,RAC}. Ref.~\cite{PR3_PRL} applied it to polarization data from the \textit{Planck} mission High-Frequency Instrument (HFI) third public release (PR3)~\cite{PR3} and obtained a birefringence measurement of $\beta=0.35^\circ\pm0.14^\circ$ ($68\%$ C.L.), with no apparent contribution from systematic uncertainties. 

The subsequent study of HFI data from \textit{Planck}'s fourth public release (known as PR4 or \texttt{NPIPE} reprocessing)~\cite{NPIPE} done in Ref.~\cite{NPIPE_PRL} yielded a birefringence angle of $\beta=0.30^\circ\pm0.11^\circ$ ($68\%$ C.L.). More importantly, that study revealed that, although robust against systematics, the methodology is sensitive to the $EB$ correlation inherent in polarized foreground emission. The contribution from a possible foreground $EB$ correlation had been considered but ultimately neglected in previous works~\cite{Minami2019_auto, Minami2020cutsky, Minami2020cross, PR3_PRL}, since the $EB$ correlation of both Galactic synchrotron and dust emissions is still statistically compatible with zero according to current experimental constraints~\cite{Planck_dust,Felice_sync}. 

Nevertheless, the misalignment between the filamentary dust structures of the interstellar medium and the plane-of-sky orientation of the Galactic magnetic field is expected to induce a non-null $EB$ correlation on Galactic dust emission that can bias the measurement of birefringence~\cite{Huffenberger2020,Clark2021,Cukierman2022}. Two independent approaches to model dust $EB$ and correct for such a bias were proposed in Ref.~\cite{NPIPE_PRL}: one based on the $EB$ correlation predicted from the misalignment of dust filaments and magnetic field lines~\cite{Clark2021}; and another one that takes the $EB$ from the foreground templates produced by Bayesian component-separation analyses that fit parametric models to CMB data such as the \texttt{Commander}\footnote{\texttt{Commander} products are available at \url{https://pla.esac.esa.int/\#maps}, and the code itself at \url{https://github.com/Cosmoglobe/Commander}.} sky model~\cite{Commander_ref4, Commander_ref5, Commander_ref1, Commander_ref2}.

Produced by a different physics, no alignment mechanism is known to induce a non-null $EB$ correlation in synchrotron radiation. The study of the synchrotron-dominated frequencies of WMAP and the Low-Frequency Instrument (LFI) of \textit{Planck} in Refs.~\cite{Felice_sync, johannes, johannes_wmap} suggests that such a hypothetical synchrotron $EB$ has little effect on the measurement of birefringence. Correcting only for dust $EB$, the combined analysis of \textit{Planck} HFI and LFI with WMAP data gave a birefringence angle of $\beta=0.342^\circ{}^{+0.094^\circ}_{-0.091^\circ}$ ($68\%$ C.L.)~\cite{johannes_wmap}.

The aim of this work is to test the robustness of these cosmic birefringence measurements against Galactic foreground emission and instrumental systematics using high-fidelity simulations of \textit{Planck} data. Such an analysis was part of the study on the impact of systematics undertaken in Ref.~\cite{NPIPE_PRL}, but finally not described in that publication due to space limitations. Although the results presented here are restricted to simulations of \textit{Planck} HFI, our conclusions on the impact of dust $EB$ and the robustness of the methodology against instrumental systematics are expected to extend to the other measurements presented above.

The original implementation of the methodology presented in Refs.~\cite{Minami2019_auto, Minami2020cutsky, Minami2020cross} relies on Markov chain Monte Carlo (MCMC) methods to sample the likelihood and obtain the posterior distribution. To reduce the computational cost of that approach, in this work, we present an iterative algorithm based on the small-angle approximation to semi-analytically calculate the maximum likelihood solution. With this implementation we achieve a great reduction of execution time without compromising accuracy and precision, making the algorithm ideal for simulation-based studies of different experimental configurations, foreground models, or systematic effects. This method is the extension to the simultaneous determination of both cosmic birefringence and miscalibrated polarization angles of the methodology originally presented in Ref.~\cite{elena_alpha}.

This work is structured as follows. In section~\ref{sec:methodology}, we present our methodology for the simultaneous estimation of birefringence and miscalibration angles. To test and validate our algorithm in a realistic scenario, we use the official end-to-end simulations provided in the \texttt{NPIPE} data release~\cite{NPIPE} to build the two simulation sets described in section~\ref{sec:simulations}. The effect that Galactic foregrounds, instrumental systematics, and instrumental noise bias have on our estimates are considered in sections \ref{sec:foregrounds}, \ref{sec:systematics}, and \ref{sec: noise bias}, respectively. Final comments and conclusions are left for section~\ref{sec:conclusions}. Some technical aspects regarding the more general formulation of the estimator in terms of frequency cross-spectra, the comparison with the standard MCMC implementation, the calculation of the covariance matrix, and the modeling of Galactic foregrounds in the covariance matrix, are presented in appendices \ref{sec:appendix cross estimator and covariance}, \ref{sec:appendix MCMC compatibility}, \ref{sec:appendix covariance matrix}, and \ref{sec:appendix foreground EB in cov}, respectively.

\section{Methodology}
\label{sec:methodology}

Both the isotropic birefringence angle $\beta$ and the $\alpha_i$ miscalibration of polarization angles rotate the polarization signal observed by CMB experiments at any given frequency band $\nu_i$. However, the amplitude of the birefringence rotation depends on the difference between the value of the pseudoscalar field at the moments of photon emission and observation. For fields that vary slowly, this means that birefringence is proportional to the propagation length of photons~\cite{Eiichiro_review, Marsh2016}. In that case, we can assume that the birefringence suffered by locally emitted Galactic foregrounds ($z\approx0$) is negligible compared to that seen by CMB photons emitted at recombination ($z\approx1100$). Thus, Galactic foreground emission would only be significantly affected by the $\alpha_i$ miscalibration, allowing us to break the degeneracy between both angles \cite{Minami2019_auto}. In this way, the $E$- and $B$- mode spherical harmonic coefficients of the observed signal at a certain frequency band $\nu_i$ would be
\begin{equation}\label{eq: spherical harmonic coefficients}
    \begin{pmatrix}
		E_{\ell m}^{i,\mathrm{o}} \\
	    B_{\ell m}^{i,\mathrm{o}}
	\end{pmatrix}
	=
	\begin{pmatrix}
		\Cos{2\alpha_i} & -\Sin{2\alpha_i} \\
	    \Sin{2\alpha_i} & \phantom{-}\Cos{2\alpha_i}
	\end{pmatrix}
	\begin{pmatrix}
		E_{\ell m}^{i,\mathrm{fg}} \\
	    B_{\ell m}^{i,\mathrm{fg}}
	\end{pmatrix} 
	+
	\begin{pmatrix}
		\Cos{2\alpha_i +2\beta} & -\Sin{ 2\alpha_i + 2\beta}\\
	    \Sin{2\alpha_i +2\beta} & \phantom{-}\Cos{ 2\alpha_i + 2\beta}
	\end{pmatrix}	
	\begin{pmatrix}
		E_{\ell m}^{i,\mathrm{CMB}} \\
	    B_{\ell m}^{i,\mathrm{CMB}}
	\end{pmatrix}\,,
\end{equation}
where the different superscripts stand for the observed signal (``o''), and the underlying Galactic foreground (``fg'') and CMB emissions. Note that in this equation, and throughout the rest of the paper unless otherwise stated, foreground and CMB spherical harmonic coefficients and angular power spectra are assumed to be convolved by the instrumental beam and pixel window functions corresponding to each frequency band. 

Calculating the angular power spectra of the spherical harmonic coefficients in Eq.~(\ref{eq: spherical harmonic coefficients}) leads to the following $EE$, $BB$, and $EB$ cross-correlations between different $i$ and $j$ frequency bands:
\begin{equation}\label{eq: EE EB BE BB correlations}
    \begin{pmatrix}
        C_{\ell}^{E_i E_j, \mathrm{o}} \\
        C_{\ell}^{E_i B_j, \mathrm{o}} \\
        C_{\ell}^{B_i E_j, \mathrm{o}} \\
        C_{\ell}^{B_i B_j, \mathrm{o}} 
    \end{pmatrix}
    = \mathbf{R}(\alpha_i, \alpha_j)
    \begin{pmatrix}
        C_{\ell}^{E_i E_j, \mathrm{fg}}\\
        C_{\ell}^{E_i B_j, \mathrm{fg}} \\
        C_{\ell}^{B_i E_j, \mathrm{fg}} \\
        C_{\ell}^{B_i B_j, \mathrm{fg}} 
    \end{pmatrix}
    + \mathbf{R}(\alpha_i+\beta, \alpha_j+\beta)
    \begin{pmatrix}
        C_{\ell}^{E_i E_j, \Lambda\mathrm{CDM}}\\
        0\\
        0\\
        C_{\ell}^{B_i B_j, \Lambda\mathrm{CDM}} 
    \end{pmatrix}\, ,
\end{equation}
where $\mathbf{R}$ is the rotation matrix
\begin{equation}
    \mathbf{R}(\theta, \theta ')=
    \begin{pmatrix}
        \Cos{2\theta}\Cos{2\theta '}  & -\Cos{2\theta}\Sin{2\theta '} & -\Sin{2\theta}\Cos{2\theta '} & \phantom{-} \Sin{2\theta}\Sin{2\theta '} \\
        \Cos{2\theta}\Sin{2\theta '}  & \phantom{-}\Cos{2\theta}\Cos{2\theta '} & -\Sin{2\theta}\Sin{2\theta '} &-\Sin{2\theta}\Cos{2\theta '} \\
        \Sin{2\theta}\Cos{2\theta '} & -\Sin{2\theta}\Sin{2\theta '} & \phantom{-}\Cos{2\theta}\Cos{2\theta '} & -\Cos{2\theta}\Sin{2\theta '} \\
        \Sin{2\theta}\Sin{2\theta '}  & \phantom{-}\Sin{2\theta}\Cos{2\theta '} & \phantom{-}\Cos{2\theta}\Sin{2\theta '}  & \phantom{-}\Cos{2\theta}\Cos{2\theta '} 
    \end{pmatrix}\, .
\end{equation}
In this work, we neglect CMB $EB$ correlations prior to $\alpha_i$ or $\beta$ rotations, since they are expected to be null in $\Lambda$CDM~\cite{lue1999}. Nevertheless, in the case of working with alternative models that grant the CMB an initial $EB$ correlation at the moment of recombination (e.g., chiral gravitational waves~\cite{lue1999,chiralGW, Fujita2022} or anisotropic inflation~\cite{anisotropic_inflation}), the corresponding $EB$ terms must be added to the equations derived from Eq.~(\ref{eq: EE EB BE BB correlations}), and a theoretical angular power spectrum must be provided for them. On the other hand, we do consider a potential intrinsic foreground $EB$ correlation even though current experimental constraints find it to still be statistically compatible with zero~\cite{Planck_dust, Felice_sync}.

Starting from Eq.~(\ref{eq: EE EB BE BB correlations}), we build a maximum likelihood estimator to simultaneously calculate $\beta$ and $\alpha_i$. Although we use the cross-spectra estimator throughout the rest of the work, in this section we adopt the simpler formulation in terms of only frequency auto-spectra ($i=j$ in Eq.~(\ref{eq: EE EB BE BB correlations})) to explain the methodology in detail. For the derivation of the more general estimator in terms of frequency cross-spectra see appendix~\ref{sec:appendix cross estimator and covariance}. Following a procedure similar to the one detailed in Refs.~\cite{Minami2019_auto, elena_alpha, johannes}, the observed $EB$ correlation is written as a rotation of the observed $EE$ and $BB$ angular power spectra, the $\Lambda$CDM prediction for the CMB $EE$ and $BB$ angular power spectra, and the foreground $EB$ signal:
\begin{multline}\label{eq: equation auto-spectra}
    C_\ell^{EB,i,\mathrm{o}} = \frac{\Tan{4\alpha_i}}{2} \left( C_\ell^{EE,i,\mathrm{o}} - C_\ell^{BB,i,\mathrm{o}}\right) \\ + \cfrac{{\cal A}}{\Cos{4\alpha_i}} C_\ell^{EB,i,\mathrm{fg}} + \cfrac{ \Sin{4\beta}}{2\Cos{4\alpha_i}} \left( C_\ell^{EE,i,\Lambda\mathrm{CDM}} - C_\ell^{BB,i,\Lambda\mathrm{CDM}}\right).
\end{multline}
Here ${\cal A}$ is introduced \emph{ad hoc} as a normalization parameter: we can set ${\cal A}=0$ to ignore the foreground $EB$ contribution, or take ${\cal A}=1$ if the true foreground emission is known.

If the foreground contribution is considered (${\cal A}\neq 0$), then Eq.~(\ref{eq: equation auto-spectra}) asks for the intrinsic foreground $EB$ correlation prior to any potential $\alpha_i$ rotation. In this work, we take the \texttt{Commander}~\cite{Commander_ref4,Commander_ref5,Commander_ref1, Commander_ref2} sky model\footnote{The foreground sky model used in this work can be found at NERSC under \url{/global/cfs/cdirs/cmb/data/planck2020/all\_data/npipe6v20\_sim/skymodel\_cache}.} derived from the analysis of an early version of \textit{Planck} PR4 data as a template for the polarized foreground emission, leaving ${\cal A}$ as a free amplitude parameter to fit alongside $\beta$ and $\alpha_i$. Here we consider a single overall amplitude and use \texttt{Commander} spectral energy distributions (SEDs) to scale the foreground template to the target frequencies. The methodology extends easily to different $\mathcal{A}_i$ amplitudes for each frequency band, at the price of increasing the number of parameters to fit. 

This approach warrants a couple of caveats. First, \texttt{Commander} does not yet provide a signal-dominated template for the foreground $EB$ correlation~\cite{Planck_2016_component_separation, Planck_2018_component_separation}. Hence the template might include some of the noise fluctuations present in \textit{Planck} data. Second, the existence of miscalibrated polarization angles, which were not considered in the SEDs assumed by \texttt{Commander} to model Galactic foreground emission, might lead to a spurious $EB$ correlation in their final foreground maps. However, we believe this effect to be minimal, since the $EB$ measured in \texttt{Commander}'s dust template does not resemble a $\Sin{4\alpha} ( C_\ell^{EE,\mathrm{fg}} - C_\ell^{BB, \mathrm{fg}})/2$ rotation. To avoid such a spurious $EB$ signal, parametric component-separation methodologies that include instrumental polarization angles in their SEDs are already being proposed \cite{elena_alpha}. Finally, the integration along the line-of-sight of the thermal emission from several dust clouds with different spectral parameters and polarization angles is not fully characterized by the single modified blackbody SED used by Commander~\cite{Tassis2015, Planck2017, Vacher_method, Pelgrims2021, Ritacco2022}. This can create spurious dust $EB$ correlations with a different frequency dependence and a strong dependence on the sky fraction and multipole range considered~\cite{Vacher_SED}. Alternative ways to model the foreground $EB$ correlation without relying on templates have been proposed in Refs.~\cite{Minami2019_auto, Minami2020cutsky, NPIPE_PRL, johannes, johannes_wmap}.

From the equality in Eq.~(\ref{eq: equation auto-spectra}), we build a Gaussian likelihood to simultaneously fit for $\beta$, $\alpha_i$, and ${\cal A}$. For a CMB experiment with a total of $N_\nu$ frequency bands, and using the $\chi_{ij\ell}^{\mathrm{s}}=C_\ell^{E_iE_j,\mathrm{s}} - C_\ell^{B_iB_j,\mathrm{s}}$ abbreviation, that log-likelihood takes the form

\begin{align}\label{eq: full likelihood auto}\allowdisplaybreaks
    -2\ln {\cal L} \supset & \sum_{i,j} \sum_{\ell,\ell'} \left[ C_\ell^{EB,i,\mathrm{o}} - \cfrac{\Tan{4\alpha_i}}{2}\chi_{ii\ell}^{\mathrm{o}} - \cfrac{{\cal A}}{\Cos{4\alpha_i}}C_\ell^{EB,i,\mathrm{fg}}  - \cfrac{\Sin{4\beta}}{2\Cos{4\alpha_i}} \chi_{ii\ell}^{\Lambda\mathrm{CDM}}\right] \times\nonumber\\
    & \phantom{\sum} \mathbf{C}_{ij\ell\ell'}^{-1}\left[  C_{\ell'}^{EB,j,\mathrm{o}} - \cfrac{\Tan{4\alpha_j}}{2}\chi_{jj\ell'}^{\mathrm{o}} -\cfrac{{\cal A}}{\Cos{4\alpha_j}}C_{\ell'}^{EB,j,\mathrm{fg}}  - \cfrac{ \Sin{4\beta}}{2\Cos{4\alpha_j}} \chi_{jj\ell'}^{\Lambda\mathrm{CDM}}\right],
\end{align}
where we are summing over all possible combinations of detector channels ($i,j=1, ..., N_\nu$) and multipoles ($\ell,\ell' \in[\ell_{\mathrm{min}},\ell_{\mathrm{max}}] $ for a total of $N_\ell = \ell_{\mathrm{max}} - \ell_{\mathrm{min}}+1$), and $\mathbf{C}_{ij\ell\ell'}$ is the covariance matrix of $N_\nu N_\ell \times N_\nu N_\ell$ dimensions. Here we use the $\supset$ symbol to remind the reader that Eq.~(\ref{eq: full likelihood auto}) does not show the full log-likelihood, since the $\ln |C_{ij\ell\ell'}|$ term is not included; although it can usually be excluded from the minimization process, in our case, the log-determinant must be taken into account because the model parameters explicitly appear in the covariance matrix. Therefore, the variation of the free parameters during minimization leads to a change in the likelihood's normalization that can bias the results if it is not correctly accounted for. As will be further discussed later in this section and in appendix \ref{sec:appendix MCMC compatibility}, the iterative algorithm we propose automatically accounts for this change, so that we do not need to explicitly consider the contribution of the log-determinant.

For each combination of $ij$ frequency bands, the corresponding $N_\ell \times N_\ell$ box of the covariance is calculated as
\begin{align}\label{eq: auto cov general expresion}\allowdisplaybreaks
    \mathbf{C}_{ij\ell\ell'} = & \mathrm{Cov}\bigg[ C_\ell^{EB,i,\mathrm{o}} - \cfrac{\Tan{4\alpha_i}}{2}\chi_{ii\ell}^{\mathrm{o}} - \cfrac{{\cal A}}{\Cos{4\alpha_i}}C_\ell^{EB,i,\mathrm{fg}} - \cfrac{ \Sin{4\beta}}{2\Cos{4\alpha_i}} \chi_{ii\ell}^{\Lambda\mathrm{CDM}},\nonumber\\
    & \phantom{\mathrm{Cov}\Big[} C_{\ell'}^{EB,j,\mathrm{o}} - \cfrac{\Tan{4\alpha_j}}{2}\chi_{jj\ell'}^{\mathrm{o}} - \cfrac{{\cal A}}{\Cos{4\alpha_j}}C_{\ell'}^{EB,j,\mathrm{fg}} - \cfrac{ \Sin{4\beta}}{2\Cos{4\alpha_j}} \chi_{jj\ell'}^{\Lambda\mathrm{CDM}}\bigg].
\end{align}
Note that in Eq.~(\ref{eq: auto cov general expresion}), covariance elements are calculated from the observed angular power spectra and from the model for both foreground and CMB signals. Neglecting $\ell$-to-$\ell'$ correlations and assuming that the spherical harmonic coefficients are Gaussian, we can approximate each box of the covariance between whatever X, Y, Z, and W combination of observed, foreground, or CMB E- and B-modes by its diagonal:
\begin{equation}\label{eq: approx no lxl'}
\mathrm{Cov}\left[ C_\ell^{XY}, C_{\ell'}^{ZW} \right]\approx \frac{1}{2\ell+1} \delta_{\ell\ell'}\left( C_\ell^{XZ}C_\ell^{YW} +C_\ell^{XW}C_\ell^{YZ} \right).
\end{equation}
In our notation, we explicitly indicate the use of this approximation by reducing $\mathbf{C}_{ij\ell\ell'}$ to $\mathbf{C}_{ij\ell}$, and summations in both $\ell$ and $\ell'$ to just $\ell$. The impact that the non-Gaussianity of Galactic foregrounds has on the estimator was already studied in Ref.~\cite{elena_alpha}. In the case of partial skies, one can still approximate the covariance matrix as diagonal as long as the $\ell$-to-$\ell'$ correlations induced by the limited sky coverage are reduced by sufficiently apodizing the analysis mask and binning the angular power spectra. See appendix~\ref{sec:appendix covariance matrix} for more details in the calculation of the covariance matrix. Finally, we average both the angular power spectra and the covariance matrix into $N_{\mathrm{bins}}$ uniform bins of $\Delta\ell$ width:
\begin{equation}
    C_b^X=\frac{1}{\Delta\ell} \sum_{\ell\in b} C_\ell^X, \hspace{3mm} \mathbf{C}_{ijb}=\frac{1}{\Delta\ell^2} \sum_{\ell\in b} \mathbf{C}_{ij\ell}.
\end{equation}
In addition to reducing the coupling between non-diagonal multipoles for masked skies, binning also helps to reduce the numerical instabilities that arise from calculating the covariance matrix from observed spectra rather than from theoretical models~\cite{elena_alpha}. In this work, we focus on high-$\ell$ data and uniformly bin angular power spectra and covariance matrices from $\ell_{\mathrm{min}}=51$ to $\ell_{\mathrm{max}}=1490$, with a spacing of $\Delta\ell=20$ ($N_{\mathrm{bins}}=72$), to match the analysis in Ref.~\cite{NPIPE_PRL}.

The likelihood in Eq.~(\ref{eq: full likelihood auto}) is often sampled with Markov chain Monte Carlo (MCMC) methods to find the best-fit solution for all parameters, as done in Refs.~\cite{Minami2019_auto, Minami2020cutsky, Minami2020cross, PR3_PRL, NPIPE_PRL, johannes, johannes_wmap}. As an extension of the methodology presented in Ref.~\cite{elena_alpha}, we propose an alternative iterative implementation to calculate the maximum likelihood solution for $\mathcal{A}$, $\beta$, and $\alpha_i$ semi-analytically. In this algorithm, we assume that the $\mathcal{A}$, $\beta$, and $\alpha_i$ parameters in the covariance matrix are known and fixed, starting at $\mathsf{x}_i = (\mathcal{A}, \beta, \alpha_i) = (1, 0, 0)$. Then, applying the small-angle approximation (valid for angles $\lesssim 10^\circ)$, the likelihood in Eq.~(\ref{eq: full likelihood auto}) is reduced to
\begin{align}\label{eq: likelihood auto small angle}\allowdisplaybreaks
    -2\ln {\cal L} \supset & \sum_{i,j} \sum_b \left[ C_b^{EB,i,\mathrm{o}} - 2\alpha_i\chi_{iib}^{\mathrm{o}} - {\cal A}C_b^{EB,i,\mathrm{fg}} - 2\beta\chi_{iib}^{\Lambda\mathrm{CDM}}\right]\times \nonumber\\
    & \phantom{\sum} \mathbf{C}_{ijb}^{-1}\left[  C_b^{EB,j,\mathrm{o}} - 2\alpha_j\chi_{jjb}^{\mathrm{o}} -{\cal A}C_b^{EB,j,\mathrm{fg}}  - 2\beta\chi_{jjb}^{\Lambda\mathrm{CDM}}\right].
\end{align}
Differentiating Eq.~(\ref{eq: likelihood auto small angle}) with respect to each of the $\mathsf{x}_i$ parameters, we obtain a set of linear equations with which to calculate the maximum likelihood solution for all parameters analytically. This first estimate is then used to update the covariance matrix and recalculate a new best-fit solution, starting an iterative process that converges after only a few iterations. By fixing the value of the free parameters in the covariance matrix and iteratively updating them, we are implicitly accounting for the change in the likelihood's normalization that would otherwise need to be explicitly considered through the inclusion of the $\ln |C_{ij\ell}|$ term in Eqs.~(\ref{eq: full likelihood auto}) and~(\ref{eq: likelihood auto small angle}). With this algorithm, we achieve a great reduction of execution time without losing accuracy and precision with respect to the MCMC sampling of the full likelihood. See appendix \ref{sec:appendix MCMC compatibility} for a more detailed comparison of both implementations.

In particular, the minimization of Eq.~(\ref{eq: likelihood auto small angle}) leads to a linear system $\sum_n \mathsf{A}_{mn} \mathsf{x}_n = \mathsf{b}_m$ of the form:
\begin{equation}\label{eq: linear system}
  \begin{pmatrix}
    \begin{array}{c|c|ccc}
        \Xi & Z & K_n & \dots \\
        \hline
        Z   & \Theta & T_n & \dots \\
        \hline
        K_m & T_m &  &  & \\
        \vdots & \vdots &  \multicolumn{3}{c}{\smash{\raisebox{.7\normalbaselineskip}{$\Omega_{mn}$}}}\\
    \end{array}
  \end{pmatrix}
    \begin{pmatrix}
        {\cal A} \\
        \beta \\
        \alpha_n  \\
        \vdots
    \end{pmatrix}\, 
    =
     \begin{pmatrix}
        \xi \\
        \theta \\
        \omega_m  \\
        \vdots
    \end{pmatrix}\,   ,
\end{equation}
where the elements of the $\mathsf{A}_{mn}$ system matrix are
\begin{align} \allowdisplaybreaks 
    \Xi =& \phantom{2} \sum_{i,j} \sum_b C_b^{EB,i,\mathrm{fg}} \mathbf{C}_{ijb}^{-1} C_b^{EB,j,\mathrm{fg}}, \label{eq: first term auto system}\\
    Z =& 2 \sum_{i,j} \sum_b  C_b^{EB,i,\mathrm{fg}} \mathbf{C}_{ijb}^{-1} \chi_{jjb}^{\Lambda\mathrm{CDM}}, \\
    \Theta =& 4 \sum_{i,j} \sum_b \chi_{iib}^{\Lambda\mathrm{CDM}}  \mathbf{C}_{ijb}^{-1} \chi_{jjb}^{\Lambda\mathrm{CDM}},
\end{align}
\begin{align} \allowdisplaybreaks 
    K_m=& 2 \sum_j \sum_b \chi_{mmb}^{\mathrm{o}}\mathbf{C}_{mjb}^{-1} C_b^{EB,j,\mathrm{fg}}, \\
    T_m=& 4 \sum_j \sum_b \chi_{mmb}^{\mathrm{o}}\mathbf{C}_{mjb}^{-1} \chi_{jjb}^{\Lambda\mathrm{CDM}}, \\
    \Omega_{mn} =& 4 \sum_b \chi_{mmb}^{\mathrm{o}} \mathbf{C}_{mnb}^{-1} \chi_{nnb}^{\mathrm{o}}, 
\end{align}
and $\mathsf{b}_m$ terms that are
\begin{align}\allowdisplaybreaks
    \xi =&  \phantom{2} \sum_{i,j} \sum_b C_b^{EB,i,\mathrm{fg}} \mathbf{C}_{ijb}^{-1} C_b^{EB,j,\mathrm{o}},  \\
    \theta =& 2 \sum_{i,j} \sum_b C_b^{EB,i,\mathrm{o}} \mathbf{C}_{ijb}^{-1} \chi_{jjb}^{\Lambda\mathrm{CDM}}, \\
    \omega_m =& 2 \sum_j \sum_b \chi_{mmb}^{\mathrm{o}} \mathbf{C}_{mjb}^{-1}  C_b^{EB,j,\mathrm{o}} . \label{eq: last term auto system}
\end{align}
Finally, this formalism allows us to calculate the uncertainty associated with the maximum likelihood solution within the Fisher matrix approximation. The corresponding covariance matrix is $\mathsf{C}^{-1}_{mn}= -\frac{\partial^2{\ln\cal L}}{\partial \mathsf{x}_m \partial \mathsf{x}_n} =\mathsf{A}_{mn}$.

\section{\texttt{NPIPE} simulations and Galactic masks}
\label{sec:simulations}

We use the official \texttt{NPIPE} end-to-end simulations\footnote{The foreground sky model and the simulations used in this paper (individual input components as well as coadded maps) are available at NERSC under \url{/global/cfs/cdirs/cmb/data/planck2020/all\_data.}} of \textit{Planck}'s HFI 100, 143, 217, and 353GHz bands to test the robustness of our methodology against Galactic foreground emission and instrumental systematics in a realistic scenario. The \texttt{NPIPE} release~\cite{NPIPE} provides a set of high-fidelity Monte Carlo simulated maps that include CMB, Galactic foregrounds, noise, and systematics. Simulations of detector splits, obtained by dividing the horns in the focal plane into two subsets (A and B) and independently processing them, are also provided. 

The CMB realizations are the full-focal plane simulations used in PR3~\cite{FFP10sims-PR3}. Galactic foregrounds are simulated by evaluating the \texttt{Commander} sky model derived from the analysis of an early version of \texttt{NPIPE} data at the target frequencies. In particular, synchrotron radiation is modeled with a power-law SED, and thermal dust emission as a one-component modified blackbody. When the angular resolution of the foreground model is higher than that of the target frequency band (e.g., dust at 100GHz), the foreground component is smoothed to match the \texttt{QuickPol}~\cite{quickpol} beam specific to the \texttt{NPIPE} dataset. To avoid divergences in the deconvolution, the Gaussian beam with full-width-at-half-maximum (FWHM) of 5 arcmin present in \texttt{Commander}’s dust component is maintained at 217 and 353GHz. Static zodiacal emission is also included by adding the same nuisance templates that \texttt{Commander} marginalized over. Among other instrumental effects, noise maps include beam systematics, gain calibration and bandpass mismatches, analogue-to-digital conversion non-linearities, and the transfer-function corrections. Noise maps also capture the non-linear response of the instrument and of the \texttt{NPIPE} processing pipeline, reproducing the non-linear couplings between signal and noise that they introduce. Please refer to Ref.~\cite{NPIPE} for a more detailed description of the systematic effects included in \texttt{NPIPE} simulations.

For the analysis presented in section~\ref{sec:foregrounds}, we build a first simulation set ($\mathtt{FG^\alpha + CMB^{\alpha+\beta}}$ $\mathtt{+N}$) by coadding foreground maps with 100 different CMB realizations and their associated noise maps. Before addition, foreground and CMB maps are rotated by $\alpha_i$ and $\alpha_i + \beta$ angles, respectively. For each realization, birefringence and polarization angles are randomly drawn from a uniform distribution in the range $[-1^\circ,1^\circ]$. To mimic the analysis in Ref.~\cite{NPIPE_PRL}, we simulate A/B detector splits with a different miscalibration angle per split, i.e., $\alpha_i$ with $i=$100A, 100B, ..., 353B. In other words, we treat A/B detector splits as if they were observations from different frequency bands.

To assess the impact of instrumental systematics different from a miscalibration of polarization angles, we build a second simulation set ($\mathtt{CMB+N}$) by coadding the same 100 CMB realizations and their associated noise maps. In this case we do not rotate the maps, since we want to use them to test whether some of the systematic effects in the \texttt{NPIPE} data lead to any systematic $\beta$ or $\alpha_i$ angles. This second simulation set is also generated for A/B detector splits.

\begin{figure}[t]
    \centering
    \includegraphics[width=0.85\linewidth]{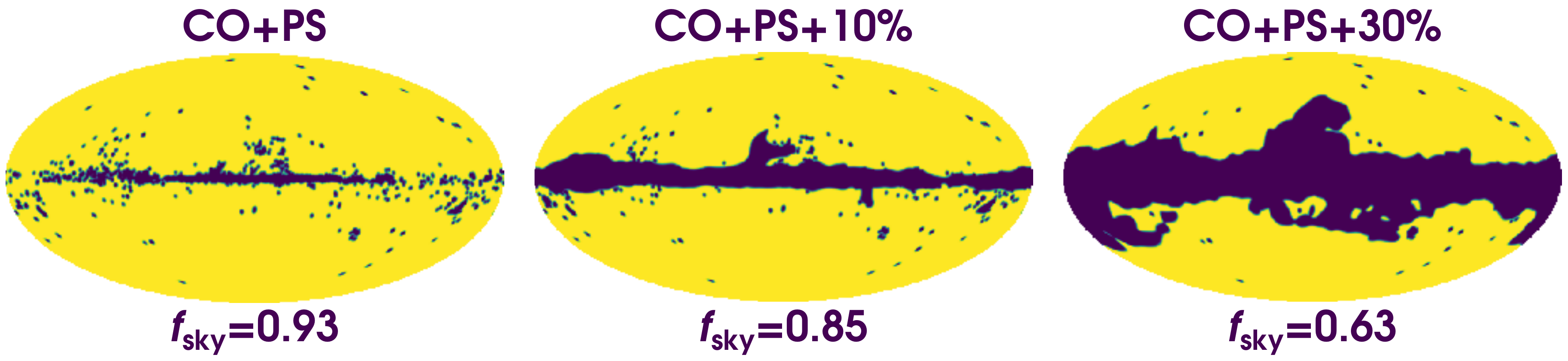}
    \caption{Galactic masks considered in this work. They are three of the five masks used in Ref.~\cite{NPIPE_PRL} for the analysis of \textit{Planck} HFI data. }
    \label{fig: masks PRL}
\end{figure}

We adopt three of the masks used in the analysis of \textit{Planck} HFI data presented in Ref.~\cite{NPIPE_PRL} (see figure~\ref{fig: masks PRL}). The default mask is built by masking point sources and the regions where the emission of the carbon monoxide (CO) line is the brightest. The common point-source mask is constructed from the combination of the \textit{Planck} point source-polarization masks\footnote{\texttt{HFI\_Mask\_PointSrc\_2048\_R2.00.fits} file at \url{https://pla.esac.esa.int/\#maps}.} at 100, 143, 217, and 353GHz. Pixels where the CO line is brighter than 45 $\mathrm{K}_{\mathrm{RJ}}\mathrm{kms}^{-1}$ are also masked because, although CO is not polarized, the mismatch of detector bandpasses creates a spurious polarization signal via intensity-to-polarization leakage. While CO strength varies over frequency, a common CO mask is adopted for all channels to simplify the analysis. This base \texttt{CO+PS} mask is then extended to exclude $10\%$ and $30\%$ of the regions of brightest Galactic foreground emission by thresholding the \texttt{NPIPE} 353GHz polarization and total intensity maps smoothed with a Gaussian beam with a FWHM of $10^\circ$. Finally, all masks are apodized with a $1^\circ$ FWHM Gaussian. The effective sky fraction to use in the calculation of the covariance matrix is given by $f_\mathrm{sky} = N_\mathrm{pix}^{-1} (\sum_i \omega_i^2)^2 / (\sum_i \omega_i^4 )$~\cite{fsky_1,fsky_2}, where $\omega_i$ is the value of the (non-integer) apodized mask, and $N_\mathrm{pix}$ is the total number of pixels. This yields $f_\mathrm{sky}=$0.93, 0.85, and 0.63 for the \texttt{CO+PS}, \texttt{CO+PS+10\%}, and \texttt{CO+PS+30\%} masks, respectively. 

In our analysis of masked skies, we calculate full-sky pseudo-$C_\ell$s using \texttt{NaMaster}\footnote{\url{https://github.com/LSSTDESC/NaMaster}}~\cite{namaster} and without performing any $E$/$B$ mode purification\footnote{We do not perform $E/B$ mode purification because the $E$-to-$B$ leakage produced by our masks of $f_\mathrm{sky}\gtrsim60\%$ is negligible at the angular scales $\ell>50$ used in our analysis.}. For the \texttt{CO+PS+30\%} mask, we bin the pseudo-$C_\ell$ calculated with \texttt{NaMaster} to reduce the $\ell$-to-$\ell'$ correlations induced by the partial sky coverage, so than we can still approximate the covariance matrix as diagonal.

\section{Impact of Galactic foregrounds}
\label{sec:foregrounds}

\begin{figure}[h]
    \begin{minipage}{.49\linewidth}
        \centering 
        \includegraphics[width=\textwidth]{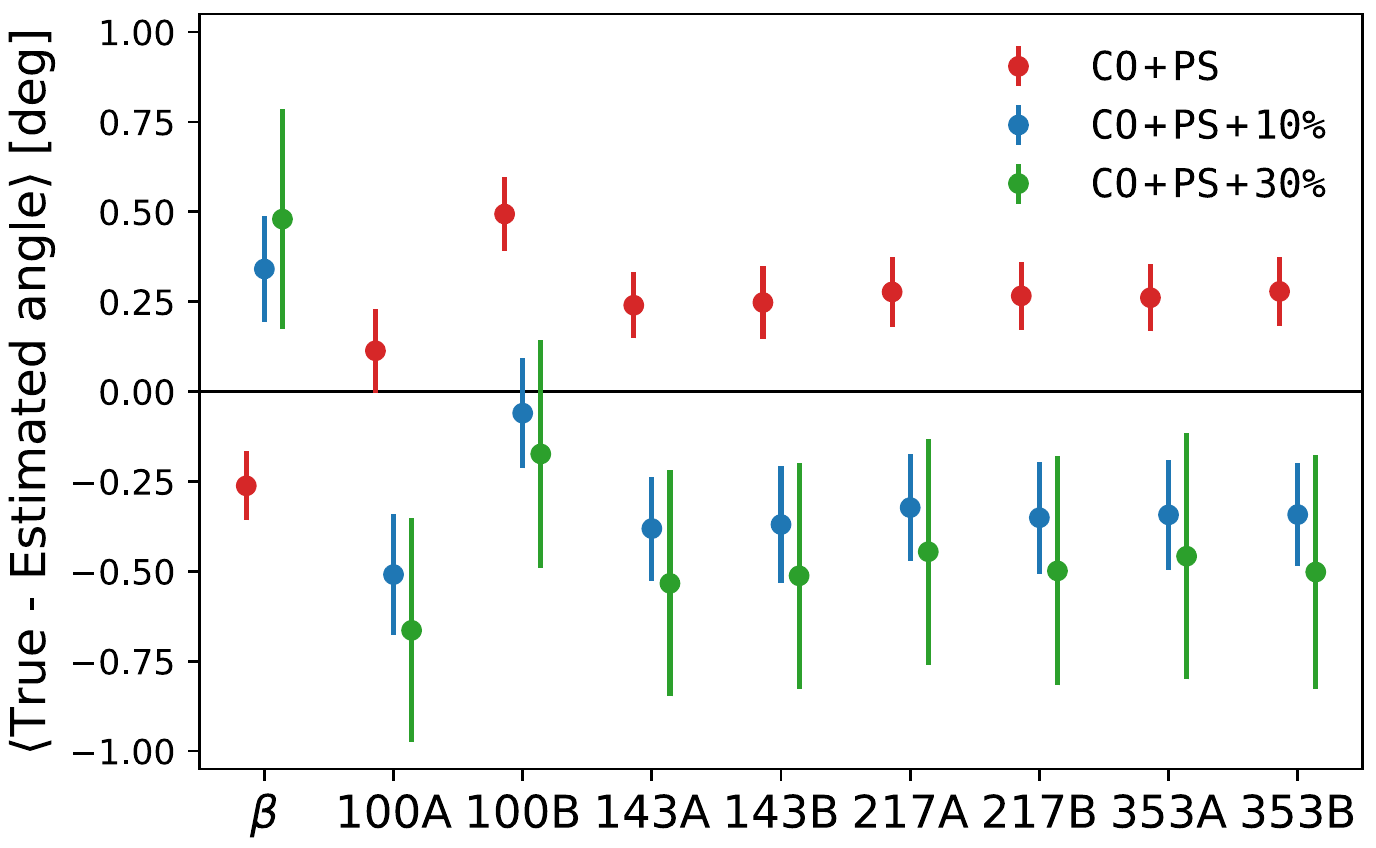}
        \caption{Bias in the simultaneous estimation of birefringence and miscalibration angles from $\mathtt{FG^\alpha + CMB^{\alpha +\beta}}$ $\mathtt{+N}$ simulations when the foreground $EB$ correlation is neglected. Uncertainties are calculated as the simulations' dispersion (one standard deviation). Results are shown for the three Galactic masks considered in this work.  \phantom{dummy text because I want an extra line here so that the figures that I have side by side are aligned }}
        \label{fig: true-estimated alpha beta angles A0}
    \end{minipage}
    \hspace{.01\textwidth}
    \begin{minipage}{.49\linewidth}
        \centering 
        \includegraphics[width=\textwidth]{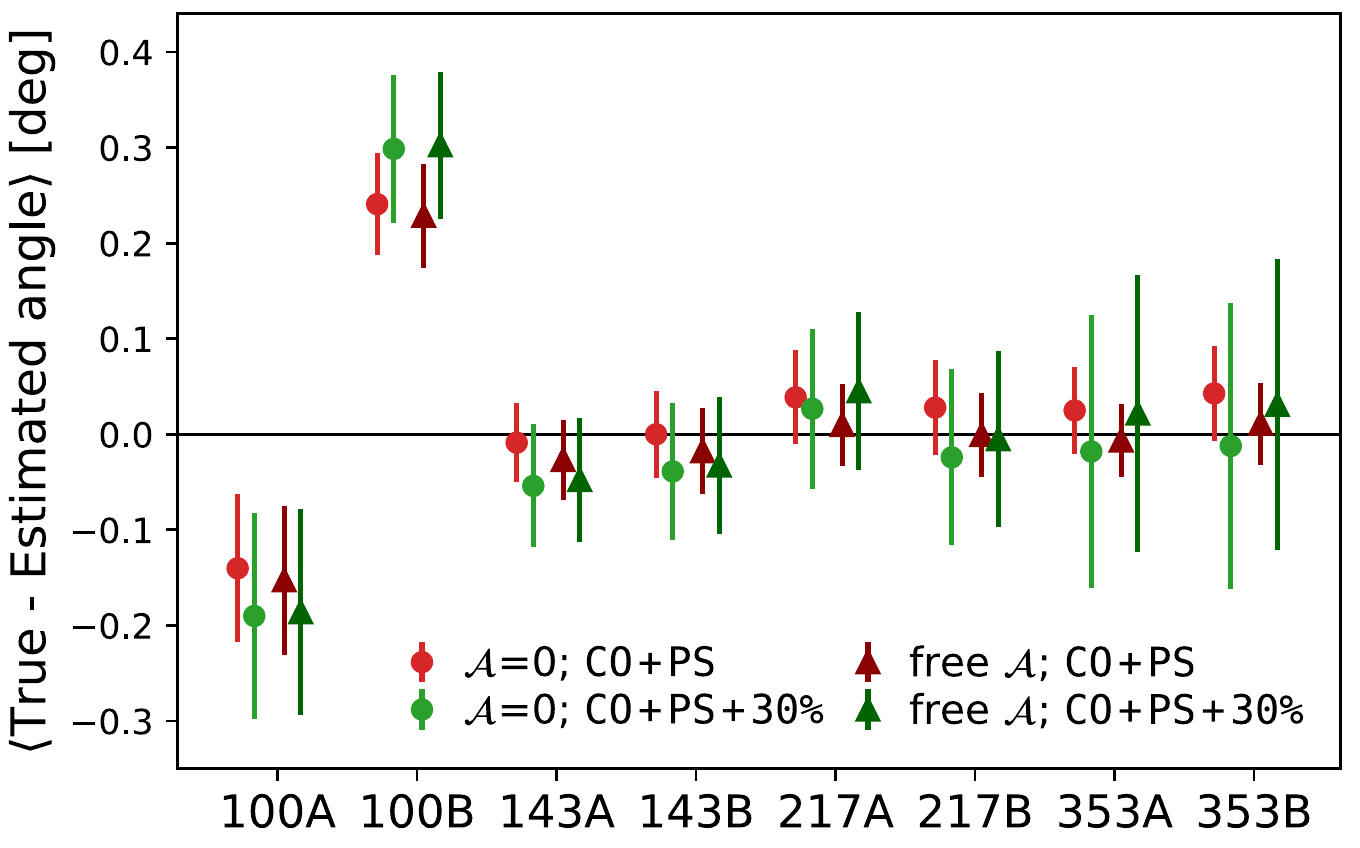}
        \caption{Bias in the estimation of exclusively miscalibration angles from $\mathtt{FG^\alpha + CMB^{\alpha +\beta}}$ $\mathtt{+N}$ simulations with $\beta=0$. We show results for the case where the foreground $EB$ correlation is neglected (circles) and considered (triangles), and only for our smallest (red) and largest (green) masks. Uncertainties are calculated as the simulations' dispersion. }
        \label{fig: bias alpha}
\end{minipage}
\end{figure}

To determine the impact of Galactic foregrounds on the measurement of cosmic birefringence, we apply our frequency cross-spectra-only estimator (see appendix~\ref{sec:appendix cross estimator and covariance}) to the 100 $\mathtt{FG^\alpha + CMB^{\alpha+\beta}+N}$ simulations and calculate the difference between the true input angles and the estimated ones. Figures \ref{fig: true-estimated alpha beta angles A0} and \ref{fig: bias alpha} show the typical bias in angle estimation, where data points correspond to the mean value and uncertainties are calculated as the simulations' dispersion (one standard deviation).

As in previous works~\cite{ Minami2019_auto, Minami2020cutsky, Minami2020cross, PR3_PRL}, we start by neglecting the foreground $EB$ contribution in figure~\ref{fig: true-estimated alpha beta angles A0}. Those results demonstrate that $\beta$ and $\alpha_i$ measurements are biased when both angles are estimated simultaneously and the foreground $EB$ is not acknowledged. In contrast, figure~\ref{fig: bias alpha} shows the typical bias that we obtain from $\mathtt{FG^\alpha + CMB^{\alpha+\beta}+N}$ simulations with $\beta=0$ when estimating miscalibration angles alone, for the case in which the foreground $EB$ is ignored ($\mathcal{A}=0$) or modeled by providing a template for foreground emission (free $\mathcal{A}$). The good agreement between the results obtained in both cases demonstrates that assuming a null foreground $EB$ correlation does not introduce any significant bias to the measurement of exclusively miscalibration angles. This was also shown by Ref.~\cite{elena_alpha} in the context of the LiteBIRD satellite. The only exceptions are the systematic $\alpha_i$ angles found for the 100A and 100B detector splits. As discussed in the next section, those systematic angles do not reflect a bias of our methodology but rather reveal the presence of a cross-polarization effect in \texttt{NPIPE} simulations.

\begin{figure}[t]
    \centering
    \includegraphics[width=\linewidth]{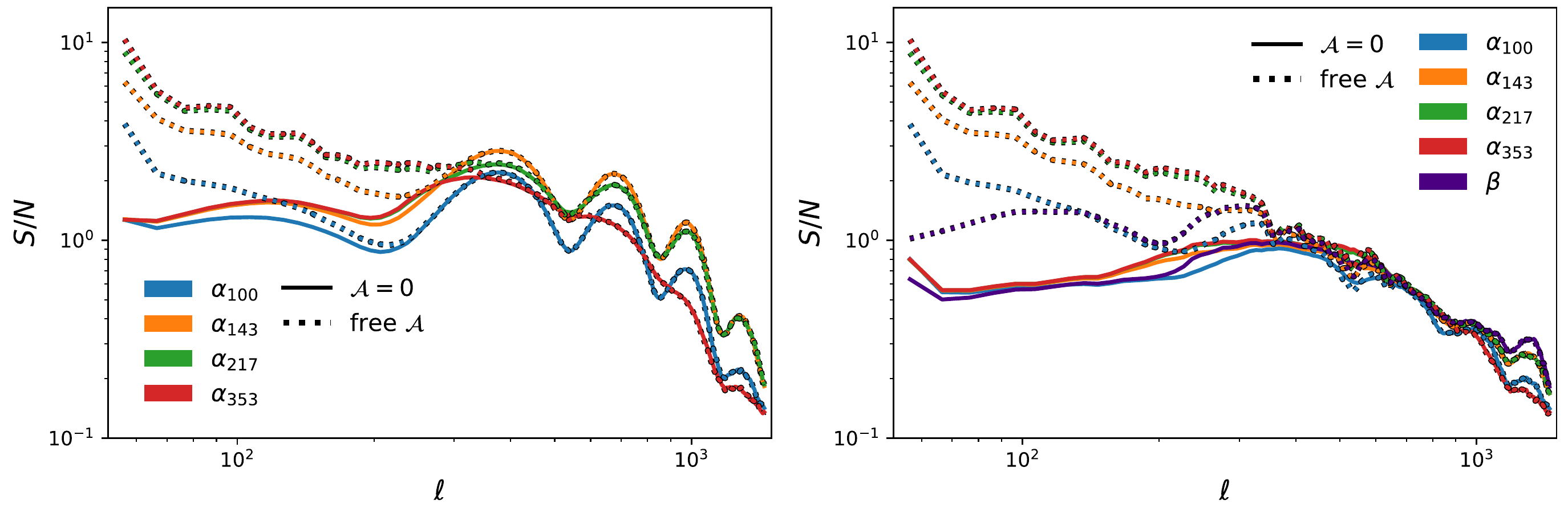}
    \caption{Signal-to-noise ratio per bin when determining exclusively miscalibration angles (left), or both birefringence and miscalibration angles simultaneously (right). Results are shown for the case where the foreground $EB$ is ignored (solid lines), or a foreground template is provided (dashed lines). The $S/N$ shown for miscalibration angles is the average of both A/B detector splits. $S/N$ ratios were calculated for one $\mathtt{FG^\alpha + CMB^{\alpha+\beta} + N}$ simulation with $\beta=\alpha_i=0.3^\circ$ and using the \texttt{CO+PS} mask.}
    \label{fig: s2n}
\end{figure}

To clarify the reason behind their different response to foreground $EB$, figure~\ref{fig: s2n} shows the signal-to-noise ratio per bin obtained for the different rotation angles with estimators that measure exclusively miscalibration angles (left), or both birefringence and miscalibration angles simultaneously (right). The signal-to-noise ratio associated with each $\mathsf{x}_m$ variable is calculated as $S/N_b(\mathsf{x}_m) = \mathsf{x}_m / (\sum_{b'\in \mathsf{w}_b} \mathsf{C}_{mmb'})^{1/2}$, where $\mathsf{w}_b$ is a square window function of $\Delta b=10$ centered around each bin, and the covariance is the $\mathsf{C}_{mn}=\mathsf{A}^{-1}_{mn}$ matrix defined in appendix~\ref{sec:appendix cross estimator and covariance}. The bin-dependence (and, by extension, $\ell$-dependence) in $\mathsf{C}_{mnb}$ comes from removing the summation in $b$ from Eqs.~(\ref{eq: first term total system}) to (\ref{eq: last term total independent term}). The main difference between both estimators is that, when focused exclusively on the determination of miscalibration angles (left panel of figure~\ref{fig: s2n}), information can be gathered from all scales, since both Galactic foregrounds and the CMB are rotated by $\alpha$. In this sense, providing a template of foreground emission increases the $S/N$ at $\ell\lesssim300$ scales, but dismissing the contribution of the foreground $EB$ correlation does not lead to a significant bias. 

\begin{figure}[t]
    \centering
    \includegraphics[width=0.95\linewidth]{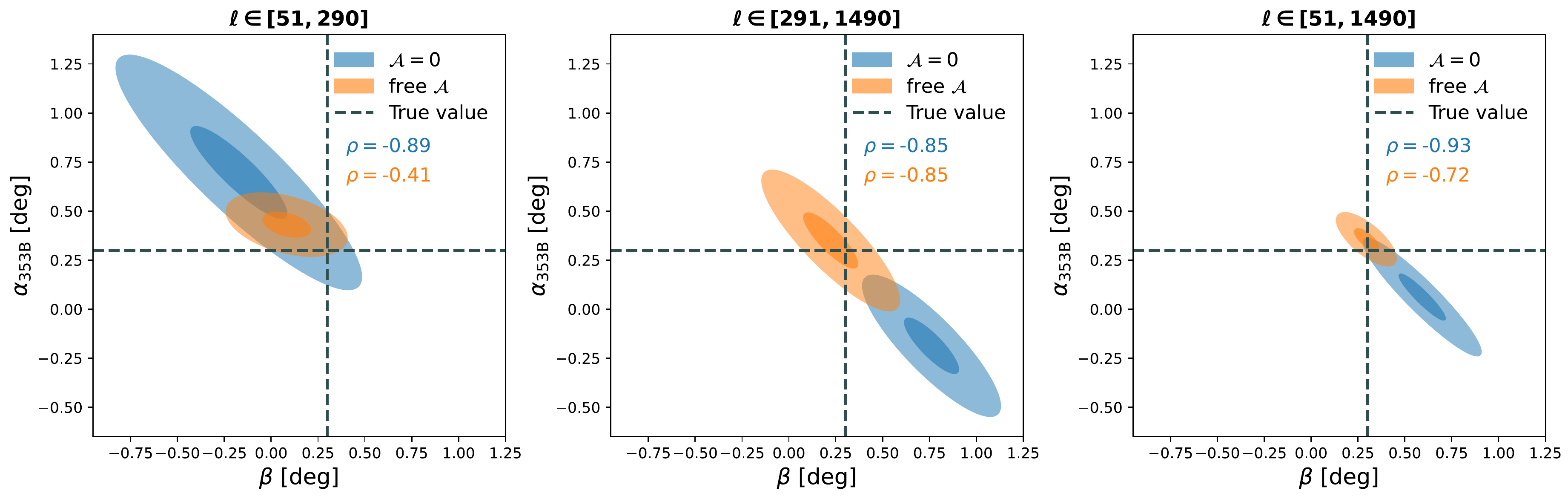}
    \caption{Correlation between $\beta$ and $\alpha_\mathrm{353B}$ angles simultaneously determined from a $\mathtt{FG^\alpha + CMB^{\alpha+\beta} + N}$ simulation with $\beta=\alpha_i=0.3^\circ$ (dashed lines) analyzed with the \texttt{CO+PS} mask. Results on the left correspond to the study of only large-scale information, those in the center to only small-scale information, and those on the right combine the information from all scales. Blue ellipses show the $1\sigma$ and $2\sigma$ Fisher confidence contours obtained when the foreground $EB$ is ignored, while the orange ones are those obtained when a foreground template is provided. Correlation coefficients are given for both cases.}
    \label{fig: ellipses}
\end{figure}

On the other hand, when trying to simultaneously determine both birefringence and miscalibration angles, we rely on foregrounds to determine $\alpha_i$ and partially break the degeneracy between both effects. Thus, a precise knowledge of foreground emission is crucial, especially at the $\ell\lesssim300$ foreground-dominated scales. In the right panel of figure~\ref{fig: s2n}, we see that when no foreground template is provided, the $S/N$ ratio of $\beta$ (purple solid line) shows the same angular dependence at large-scales as that of miscalibration angles (rest of colored solid lines), indicating that $\beta$ is being derived from foregrounds as well as the CMB. As a consequence, the unaccounted foreground $EB$ produces the bias seen in the left panel of figure~\ref{fig: ellipses} (blue contours), where we show the correlation between the $\beta$ and $\alpha_\mathrm{353B}$\footnote{Here we chose $\alpha_{\mathrm{353B}}$ as an example of a foreground-dominated band, but similar correlations are found across the rest of the detector splits.} angles recovered when only $\ell<300$ scales are used. After a template for the foreground $EB$ is provided, we see that such a bias is reduced (orange contours), and that now the angular dependence of the $S/N$ ratios of $\beta$ and $\alpha_i$ angles (dotted colored lines) correctly resemble those of, respectively, the CMB and Galactic foreground signals in the right panel of figure~\ref{fig: s2n}. In addition, the extra knowledge on foreground emission provided by the template helps to break the degeneracy between $\beta$ and $\alpha_i$ angles, relaxing the anti-correlation between them from $\rho\approx-0.9$ to $\rho\approx-0.40$.
 
At small angular scales ($\ell\gtrsim300$), the CMB starts to dominate over the foreground emission and becomes the common source of $S/N$ for both $\beta$ and $\alpha_i$. At those scales, there is not enough foreground signal to break the degeneracy between both angles, but the inclusion of the template still helps to avoid the bias induced by the foreground $EB$ correlation (central panel of figure~\ref{fig: ellipses}). Once all scales are included in the analysis (right panel of figure~\ref{fig: ellipses}), the extra constraining power that the template grants at large-scales helps to alleviate the degeneracy between both angles (from $\rho\approx-0.90$ to $\rho\approx-0.70$) and correct the bias induced by the foreground $EB$ correlation, bringing the best-fit value closer to the correct answer. In this way, ignoring the foreground $EB$ correlation when simultaneously estimating $\beta$ and $\alpha_i$ angles leads to the biases seen in figure~\ref{fig: true-estimated alpha beta angles A0}.

\begin{figure}[b]
    \centering
    \includegraphics[width=0.48\linewidth]{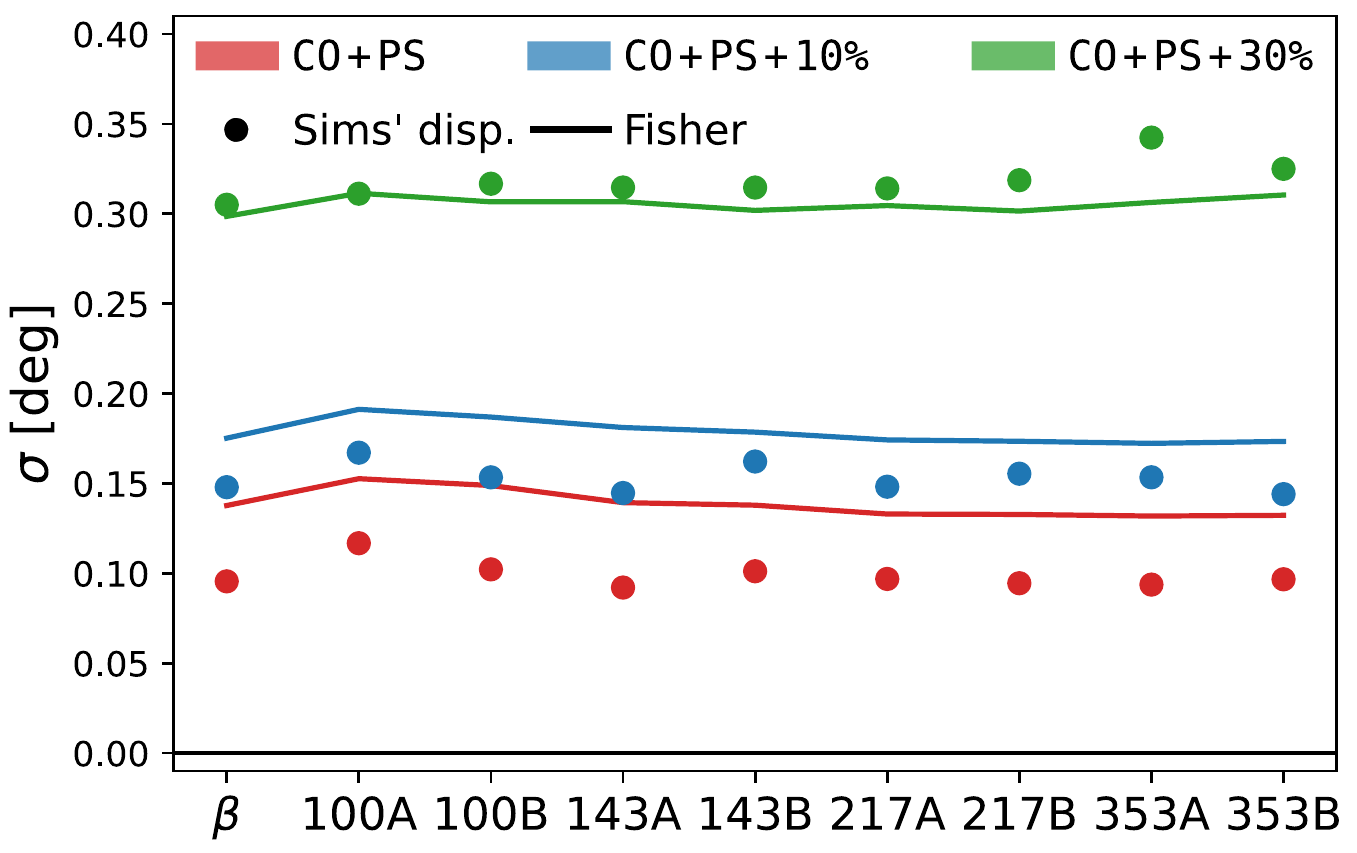}
    \caption{ Comparison between simulations- (circles) and Fisher-derived (solid lines) uncertainties when birefringence and miscalibration angles are simultaneously estimated from $\mathtt{FG^\alpha + CMB^{\alpha +\beta}}$ $\mathtt{+N}$ simulations and the foreground $EB$ correlation is neglected.}
    \label{fig: sim disp vs fisher A0}
\end{figure}

We also check the correct performance of the estimator by comparing the uncertainty obtained from the simulations' dispersion with that of the Fisher prediction. That is what we do in figure~\ref{fig: sim disp vs fisher A0}, where data points show the uncertainty calculated as the simulations' dispersion ($\sigma_{\mathrm{sim}}$) and solid lines correspond to the Fisher matrix prediction ($\sigma_{\mathrm{Fisher}}$). We expect the Fisher formalism to overestimate $\sigma_{\mathrm{sim}}$ uncertainties as seen for the \texttt{CO+PS} (red) and \texttt{CO+PS+10\%} (blue) masks, since foreground emission is a source of cosmic variance in our covariance matrix but we have used a common foreground realization for all of the $\mathtt{FG^\alpha + CMB^{\alpha+\beta}+N}$ simulations. When the majority of Galactic emission is removed from the covariance with the \texttt{CO+PS+30\%} mask (green), $\sigma_{\mathrm{sim}}$ and $\sigma_{\mathrm{Fisher}}$ uncertainties do agree.

\begin{figure}[h]
    \begin{minipage}{.49\linewidth}
        \centering 
        \includegraphics[width=1.01\textwidth]{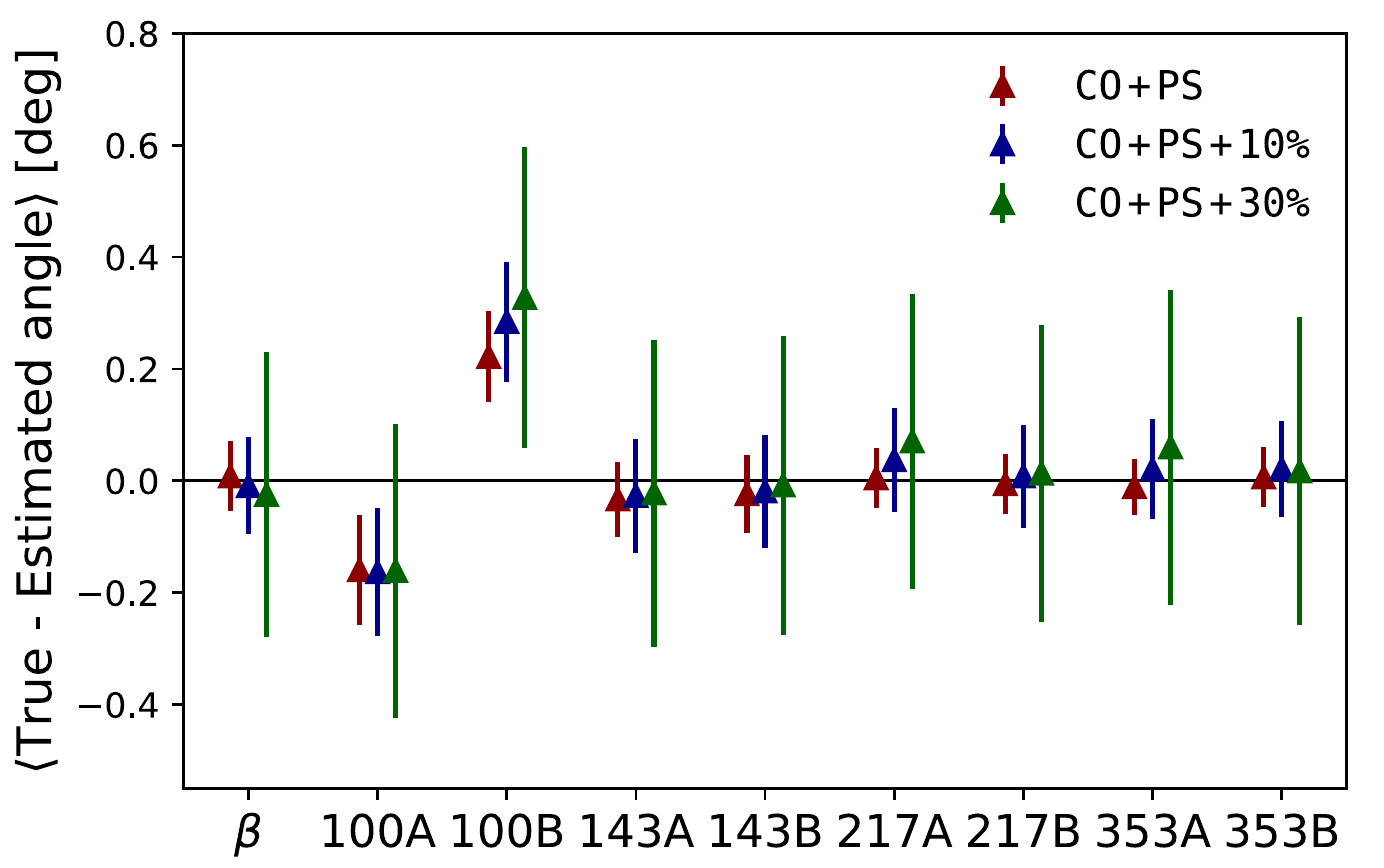}
        \caption{Bias in the simultaneous estimation of birefringence and miscalibration angles from $\mathtt{FG^\alpha + CMB^{\alpha +\beta}}$ $\mathtt{+N}$ simulations when a template for foreground emission is provided. Uncertainties are calculated as the simulations' dispersion (one standard deviation).}
        \label{fig: true-estimated alpha beta angles Afree}
    \end{minipage}
    \hspace{.01\textwidth}
    \begin{minipage}{.49\linewidth}
        \centering 
        \includegraphics[width=\textwidth]{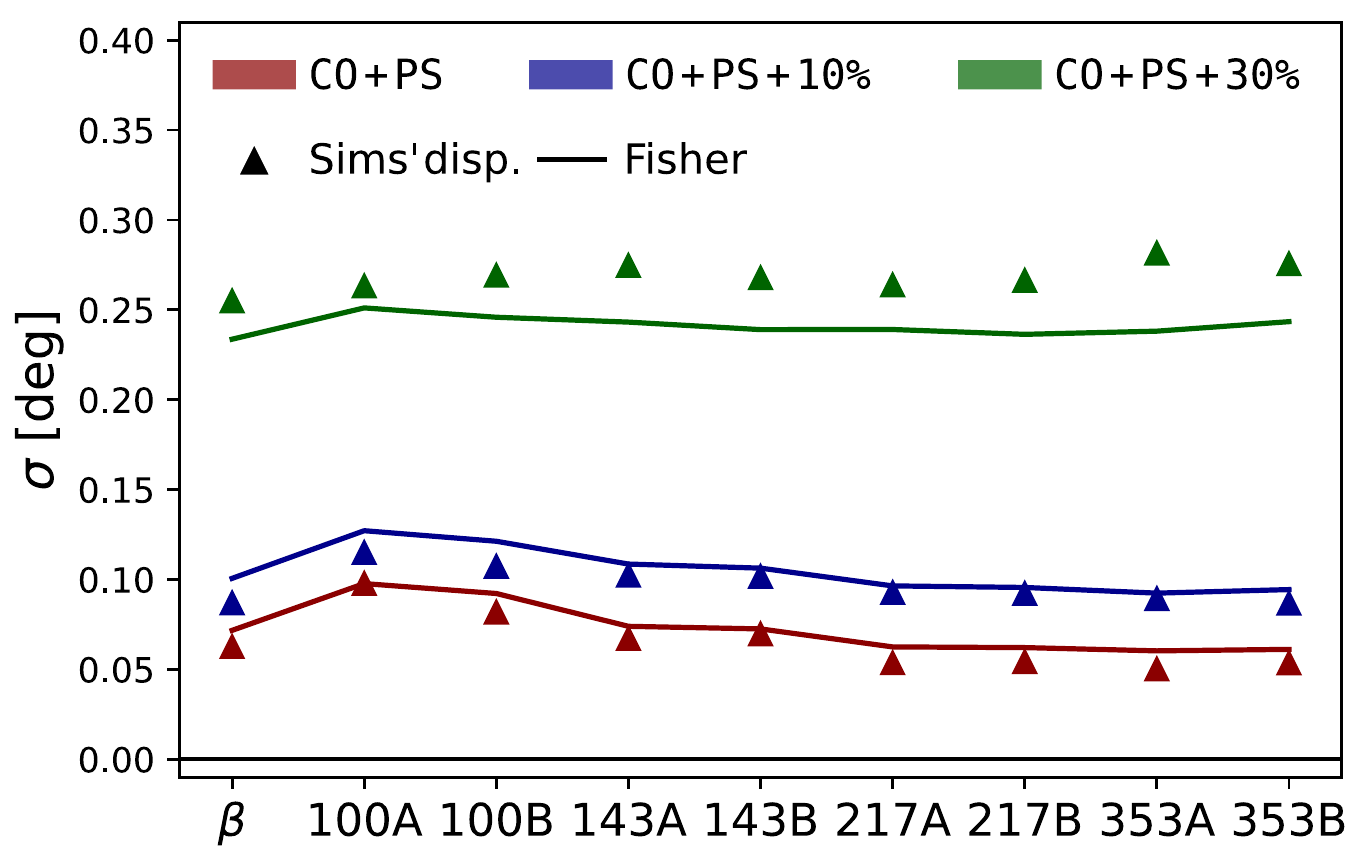}
        \caption{Comparison between simulations- (triangles) and Fisher-derived (solid lines) uncertainties when birefringence and miscalibration angles are simultaneously estimated from $\mathtt{FG^\alpha + CMB^{\alpha +\beta}}$ $\mathtt{+N}$ simulations and a foreground template is provided. }
        \label{fig: sim disp vs fisher Afree}
\end{minipage}
\end{figure}

Both the biases in the estimation of $\beta$ and $\alpha_i$ angles and the inconsistencies between simulations- and Fisher-derived uncertainties are corrected when an accurate template for foreground emission is provided and $\mathcal{A}$ is left as a free amplitude parameter in the likelihood. Figure~\ref{fig: true-estimated alpha beta angles Afree} shows that now the mean values of the recovered $\beta$ and $\alpha_i$ are centered around zero, with the exception of the aforementioned $\alpha_\mathrm{100A}$ and $\alpha_\mathrm{100B}$ systematic angles. Moreover, as is discussed in appendix~\ref{sec:appendix foreground EB in cov}, providing a template for foreground emission allows for the removal of most of the variance originated by the foregrounds' fluctuations. The effects of removing the contribution of foreground emission from the covariance are twofold. First, it leads to a reduction of the total covariance that explains the smaller uncertainties seen in figure~\ref{fig: sim disp vs fisher Afree} with respect to those in figure~\ref{fig: sim disp vs fisher A0}. And second, it ensures that the uncertainties estimated from the simulations' dispersion and the Fisher analysis are compatible with each other for all of the three Galactic masks. This last observation does not just apply to our simulations with a fixed foregrounds realization. It is a feature transferable to the analysis of real data. If we believe that our template is a measurement of the true foreground signal in the sky, then its cosmic variance should not contribute to the total uncertainty. Within this interpretation, the emission of our Galaxy is explicitly characterized at the map level through the template, while the CMB signal is only statistically characterized by the theoretical angular power spectra provided.

\begin{figure}[t]
    \centering
    \includegraphics[width=0.47\linewidth]{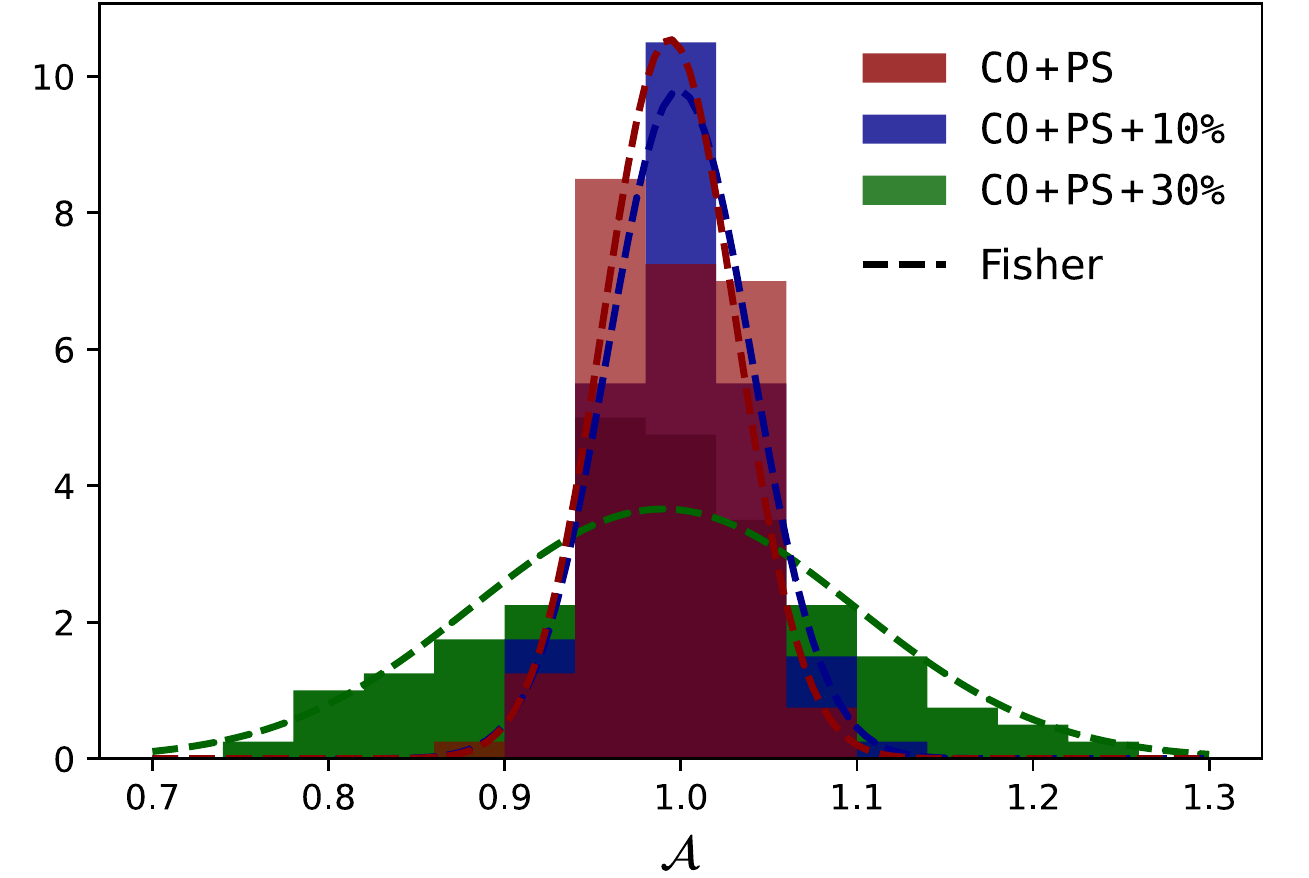}
    \caption{ Distribution of template amplitudes recovered from $\mathtt{FG^\alpha + CMB^{\alpha +\beta}}$ $\mathtt{+N}$ simulations. For comparison, dashed lines show the Gaussian distribution predicted by the Fisher analysis.}
    \label{fig: free amplitude parameter}
\end{figure}

The amplitude of the foreground template is also correctly recovered, with uncertainties from Fisher analysis and the simulations' dispersion nicely matching, as can be seen in figure~\ref{fig: free amplitude parameter}. In our case, the recovered amplitudes are centered around unity because the foreground template is the same as the fiducial foreground model used in the simulations. We checked that the choice of initial value for $\mathcal{A}$ does not condition the final results. Here we started from $\mathcal{A}=1$, but the algorithm quickly converges to compatible results after a couple more iterations when starting from $\mathcal{A}\in\{-1,0\}$. These results show that, with the exception of the systematic $\alpha_\mathrm{100A}$ and $\alpha_\mathrm{100B}$ angles, our methodology provides an unbiased estimation of both birefringence and polarization angles, once the foreground $EB$ is taken into account.

\begin{figure}[h!]
    \centering
    \includegraphics[width=0.54\linewidth]{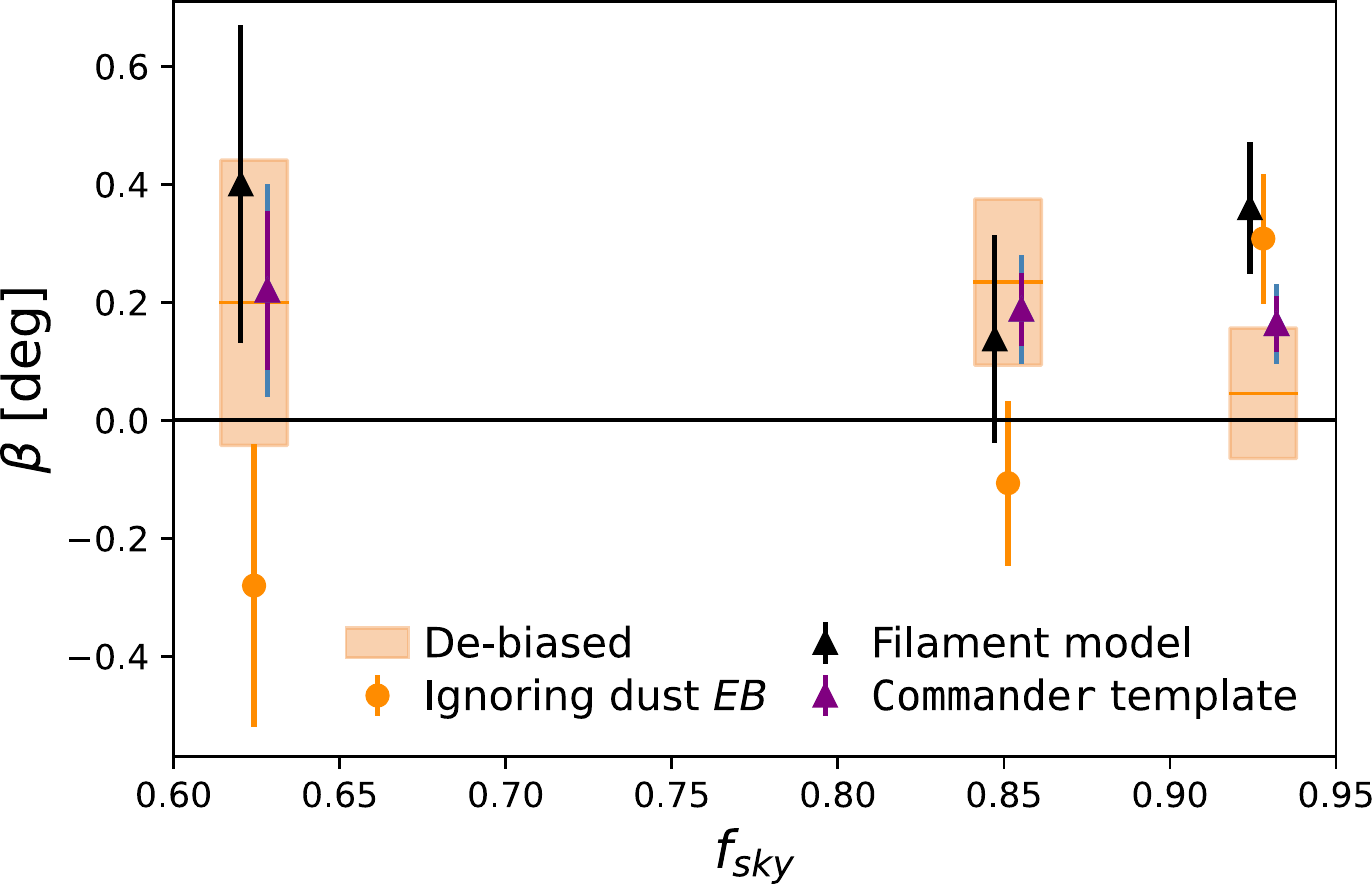}
    \caption{Birefringence measurements obtained from \textit{Planck} HFI data without accounting for the $EB$ correlation of Galactic dust, before (orange circles) and after (shaded orange squares), a posteriori correcting them with the dust $EB$ bias estimated from $\mathtt{FG^\alpha+CMB^{\alpha+\beta}+N}$ simulations. We also show the measurements obtained when dust $EB$ is modeled using either the filament model presented in Refs.~\cite{NPIPE_PRL, johannes} (black triangles) or \texttt{Commander}'s dust template (purple triangles). Gray error bars around measurements obtained with the \texttt{Commander} template show the uncertainty expected from the simulation study, while purple error bars show the actual uncertainty from the fit to \textit{Planck} data.}
    \label{fig: data}
\end{figure}

Finally, we can use the insight gained from this study of realistic simulations to interpret the results obtained from the analysis of \textit{Planck} HFI data made in Ref.~\cite{NPIPE_PRL}. For that purpose, figure~\ref{fig: data} reproduces some of the results of that publication, including the birefringence measurement obtained from \textit{Planck} data without accounting for the $EB$ correlation of Galactic dust (orange circles), and those obtained when correcting for dust $EB$ using either the \texttt{Commander} sky model (purple triangles) or the filament model presented in  Refs.~\cite{NPIPE_PRL, johannes} (black triangles). Uncertainties are calculated within the Fisher approximation. The results for the \texttt{CO+PS+30\%} mask ($f_\mathrm{sky}=0.63$) differ\footnote{The differences in the $\beta$ angle measured with and without binning are of the order of $\Delta\beta\approx0.3\sigma$ when dust $EB$ is ignored or corrected with the filament model, and of $\Delta\beta\approx0.6\sigma$ when corrected with the \texttt{Commander} sky model.}  from those reported in Ref.~\cite{NPIPE_PRL}, since now we have binned the pseudo-$C_\ell$s calculated for this mask to further reduce $\ell$-to-$\ell'$ correlations and have a more diagonal covariance matrix. 

When dust $EB$ is ignored, the decreasing values of $\beta$ found as we enlarge the Galactic mask seem to qualitatively agree with the biases expected from figure~\ref{fig: true-estimated alpha beta angles A0}. Having statistically characterized the bias produced by dust $EB$, we can de-bias those measurements by adding to them the mean bias calculated from the $\mathtt{FG^\alpha + CMB^{\alpha +\beta}}$ $\mathtt{+N}$ simulations. This leads to the shaded orange squares, which are centered at the de-biased measurements and contain all values compatible with them at $1\sigma$. De-biased values are compatible with the results obtained with both the filament and \texttt{Commander} models for $f_\mathrm{sky}<0.90$. The disagreement seen at higher $f_{\mathrm{sky}}$ suggests that the \texttt{Commander} template might be struggling to reproduce dust emission near the center of the Galactic plane where the single modified blackbody model may be too simplistic~\cite{McBride2022,Vacher_litebird,Vacher_method, Ritacco2022, Vacher_SED}.

We also find that the reduction of uncertainties that we achieve by including the foreground template in the analysis of \textit{Planck} data is larger than expected from the simulation study. To illustrate this discrepancy, the gray error bars around the birefringence measurements obtained with the \texttt{Commander} template in figure~\ref{fig: data} show the uncertainty expected from the reduction seen in the simulation study, while the purple error bars show the actual uncertainty obtained in the fit to \textit{Planck} data. In particular, uncertainties are underestimated by approximately a $18\%$, $23\%$, and $28\%$ at $f_\mathrm{sky}=$0.93, 0.85, and 0.63, respectively. The fact that uncertainties are smaller in the analysis of the data than in the analysis of simulations where \texttt{Commander} is the fiducial foreground model suggests that the template might reproduce not only foreground emission but also some of the statistical fluctuations and noise from \textit{Planck} data. In this way, the limited signal-to-noise of the \texttt{Commander} template leads to the over-reduction of the covariance matrix and the subsequent underestimation of error bars. Future experiments such as LiteBIRD~\cite{LiteBIRD_PTEP} will provide high-precision measurements of the CMB polarization that will allow us to derive a signal-dominated dust template on the full-sky.

\section{Impact of instrumental systematics}
\label{sec:systematics}

The miscalibration of polarization angles is not the only instrumental effect that interferes with the measurement of cosmic birefringence. Systematic effects like intensity-to-polarization leakage, beam leakage, or cross-polarization effects also produce spurious $EB$ correlations that can bias our analysis. Since the effect of miscalibration angles and Galactic foregrounds was already determined in the previous section, here we use $\mathtt{CMB+N}$ simulations to focus on the impact of the rest of systematics. 

By construction, $\mathtt{CMB+N}$ simulations reproduce the non-linear response of the instrument and the \texttt{NPIPE} processing pipeline, including the systematics produced by the non-linear couplings between signal and noise~\cite{NPIPE}. Therefore, $\mathtt{CMB+N}$ simulations retain the systematics associated with foregrounds (e.g., the intensity-to-polarization leakage induced by the CO bandpass mismatch), despite discarding foreground emission itself. However, without foregrounds, we are no longer able to break the degeneracy between birefringence and miscalibration angles. Hence, instead of fitting them simultaneously, we must fit for $\beta$ and $\alpha_i$ independently: we can fit a different angle for each detector split (see Ref.~\cite{elena_alpha}), knowing that these effective $\alpha_i$ yield $\alpha_i+\beta$; or we can fit the same angle for all frequency bands (see Eq.~(\ref{eq:observed_EB_CMB})), obtaining an effective birefringence angle $\beta+\bar{\alpha}$ that includes the weighted average of miscalibration angles across all detector splits. Since $\mathtt{CMB+N}$ simulations do not contain birefringence or miscalibration angles, we know that any effective $\alpha_i$ found in them are produced by the rest of the systematic effects included in the simulations, with $\bar{\alpha}$ being the net effect of those systematics in the measurement of birefringence. We refer to these angles as $\alpha_i^\mathrm{sys}$ and $\bar{\alpha}_\mathrm{sys}$.

\begin{figure}[t]
    \begin{minipage}{.49\linewidth}
        \centering 
        \includegraphics[width=\textwidth]{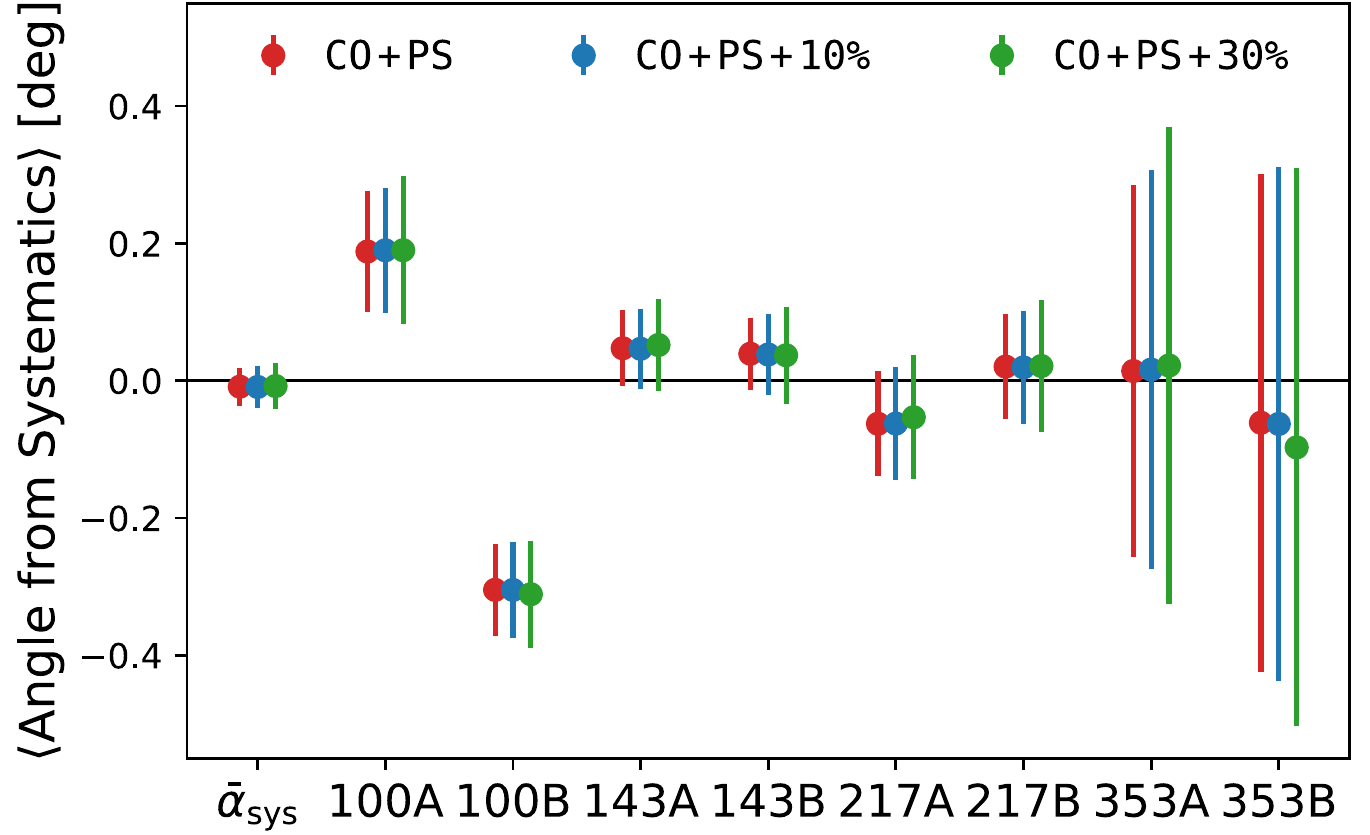}
        \caption{\label{fig: sys angles bias} Mean $\alpha_i^\mathrm{sys}$ and $\bar{\alpha}_\mathrm{sys}$ angles found in $\mathtt{CMB+N}$ simulations with the cross-spectra-only estimator. Uncertainties are calculated as the simulations' dispersion. }
    \end{minipage}
    \hspace{.01\textwidth}
    \begin{minipage}{.49\linewidth}
        \centering 
        \includegraphics[width=\textwidth]{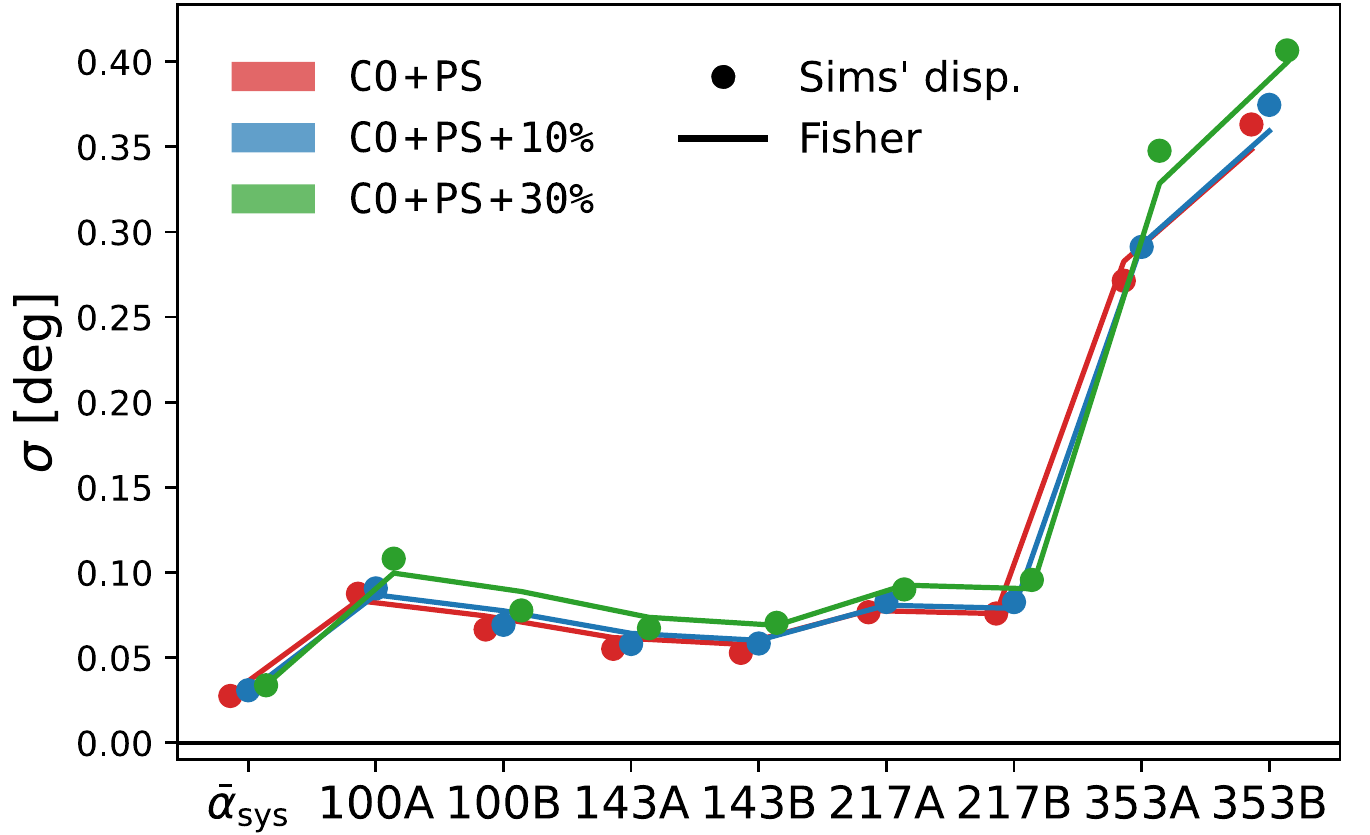}
        \caption{\label{fig: sys angles disp} Comparison between the uncertainties derived from the dispersion of $\mathtt{CMB+N}$ simulations (points) and the Fisher analysis (solid lines). }
\end{minipage}
\end{figure}

Fitting $\alpha_i^\mathrm{sys}$ and $\bar{\alpha}_\mathrm{sys}$ angles to the 100 $\mathtt{CMB+N}$ simulations with our frequency cross-spectra-only estimator, we obtain the mean angles shown in figure~\ref{fig: sys angles bias}. Their corresponding uncertainties, calculated both as the simulations' dispersion and within the Fisher approximation, are shown in figure~\ref{fig: sys angles disp}. At a first glance, uncertainties now rapidly increase at 353GHz, a behaviour that differs from the one seen in figure~\ref{fig: sim disp vs fisher Afree}. Such difference is explained by the absence of foregrounds in $\mathtt{CMB+N}$ simulations. Without foregrounds, rotation angles are estimated from the CMB, with instrumental noise as the only impediment. In this scenario, the larger uncertainties at 353GHz just reflect the configuration of \textit{Planck}-HFI. 100, 143, and 217GHz frequency bands have similar noise levels around $1.5\mu$K·deg, while the 353GHz band has $7.3\mu$K·deg~\cite{Planck_2018_overview}. Accordingly, the uncertainties recovered for 353B are approximately 6 times higher than those at, e.g., 143B, matching the roughly 6 times higher nominal noise level at 353GHz. In addition, masking the Galactic plane does not have such a dramatic effect as in figure~\ref{fig: sim disp vs fisher Afree}, because here we are fitting effective $\alpha_i+\beta$ angles instead of using foregrounds to break the degeneracy between $\alpha_i$ and $\beta$. Still, uncertainties do scale as, roughly, $f_\mathrm{sky}^{-1/2}$.

More quantitatively, we find that \texttt{NPIPE} systematics produce angles $\langle \alpha_\mathrm{100A}^\mathrm{sys} \rangle = 0.188^\circ \pm 0.009^\circ$ and $\langle \alpha_\mathrm{100B}^\mathrm{sys} \rangle = -0.305^\circ \pm 0.007^\circ$, with uncertainties given as the error of the mean. Although the values of $\alpha_i^\mathrm{sys}$ are determined to high precision using $\mathtt{CMB+N}$ simulations, we would only be able to detect them at a $1.9$-$3.8\sigma$ confidence level when simultaneously fitting $\beta$ and $\alpha_i$ to real data (compare with the uncertainties on figure~\ref{fig: sim disp vs fisher Afree}), and that is assuming that \textit{Planck}'s polarimeters were perfectly calibrated. At other frequencies, $\langle\alpha_{\mathrm{143A}}^\mathrm{sys}\rangle = 0.047^\circ \pm 0.006^\circ$, $\langle\alpha_{\mathrm{143B}}^\mathrm{sys}\rangle = 0.039^\circ \pm 0.005^\circ$, and $\langle\alpha_{\mathrm{217A}}^\mathrm{sys}\rangle = -0.063^\circ \pm 0.008^\circ$ angles are also found at a lower significance level ($0.6$-$1.3\sigma$ compared to uncertainties on figure~\ref{fig: sim disp vs fisher Afree}). These angles, produced by systematics, explain the biases seen in figures \ref{fig: true-estimated alpha beta angles A0}, \ref{fig: bias alpha}, and \ref{fig: true-estimated alpha beta angles Afree}. Note the change of sign, since those figures approximately show $-\langle \alpha_i^\mathrm{sys} \rangle$. 

\begin{figure}[t]
    \centering
    \includegraphics[width=0.96\linewidth]{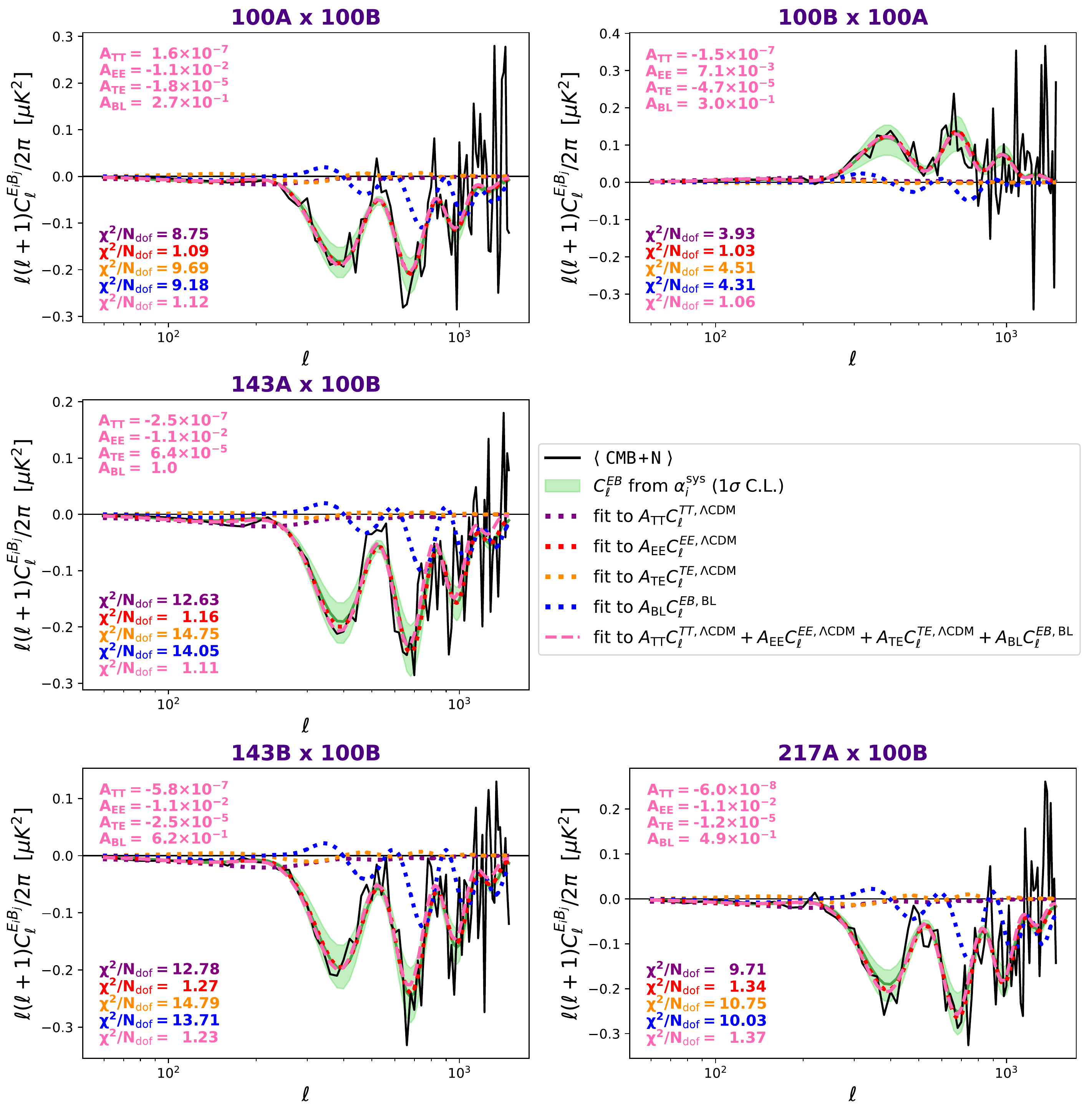}
    \caption{$EB$ angular power spectra of \texttt{CMB+N} simulations (black solid lines) obtained from the cross-correlation of a selection of detector splits where cross-polarization is found to be the main systematic effect. Spectra were calculated using the \texttt{CO+PS} mask, averaged over 100 simulations, and binned in the range $\ell \in [51,1490]$ with $\Delta\ell=20$. Superimposed are the best fits to the different $M_b$ models considered (dashed colored lines). The goodness of fit of each model is quantified by the reduced $\chi^2$ shown in the bottom-left corner of every graph, and the values of the $A_\mathrm{TT}$, $A_\mathrm{EE}$, $A_\mathrm{TE}$, and $A_\mathrm{BL}$ amplitudes obtained from the fit to Eq.~(\ref{eq: model combination all sys}) are specified in the top-left corner. Green shaded regions illustrate the $1\sigma$ confidence contours of the $EB$ correlation expected from the $\alpha_i^\mathrm{sys}$ angles found in figure~\ref{fig: sys angles bias}.}
    \label{fig: sys cross pol}
\end{figure}

\begin{figure}[t]
    \centering
    \includegraphics[width=0.96\linewidth]{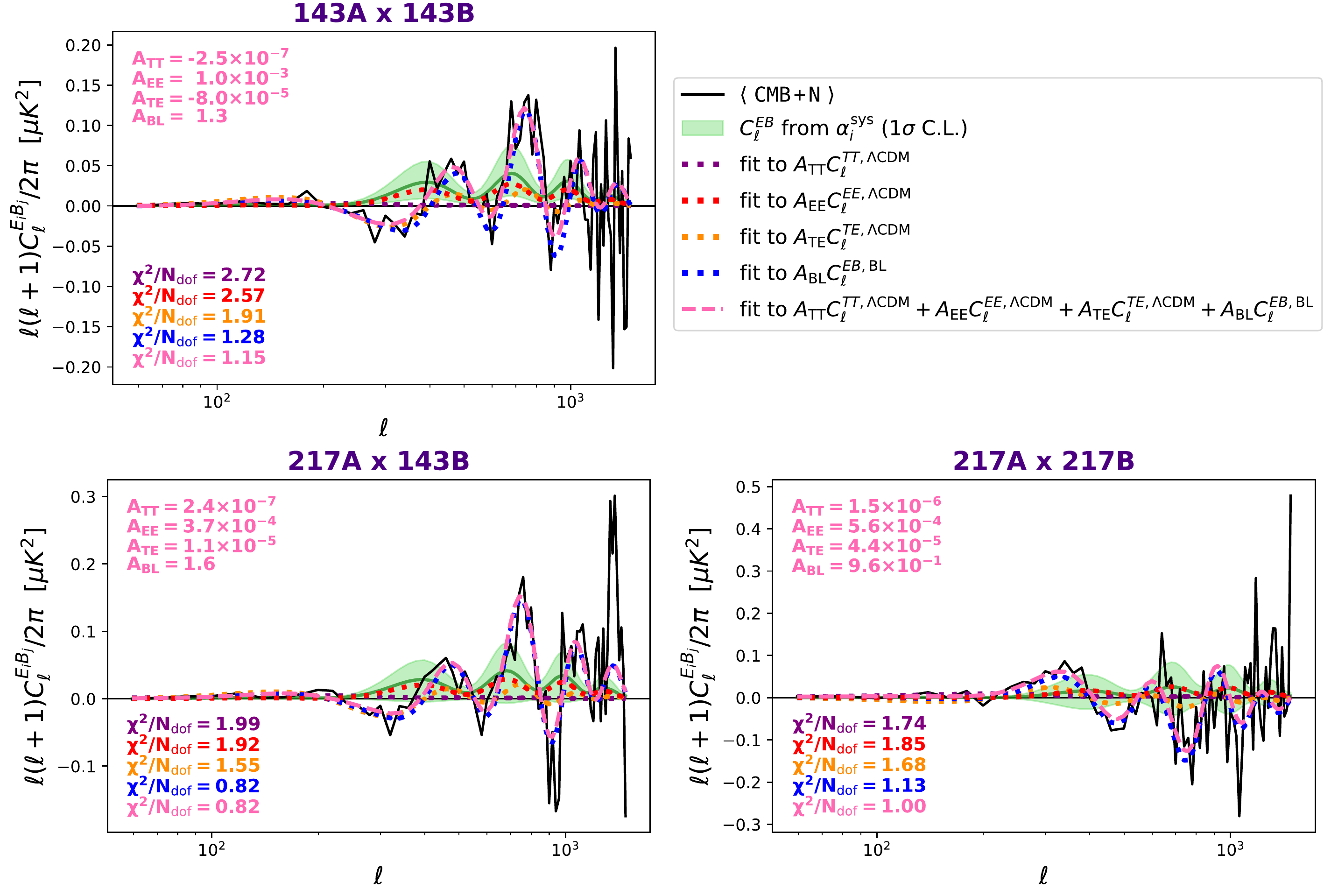}
    \caption{ Same as figure~\ref{fig: sys cross pol} but for a selection of detector splits where beam leakage is found to be the main systematic effect.}
    \label{fig: sys beam leak}
\end{figure}

\begin{figure}[t]
    \centering
    \includegraphics[width=0.96\linewidth]{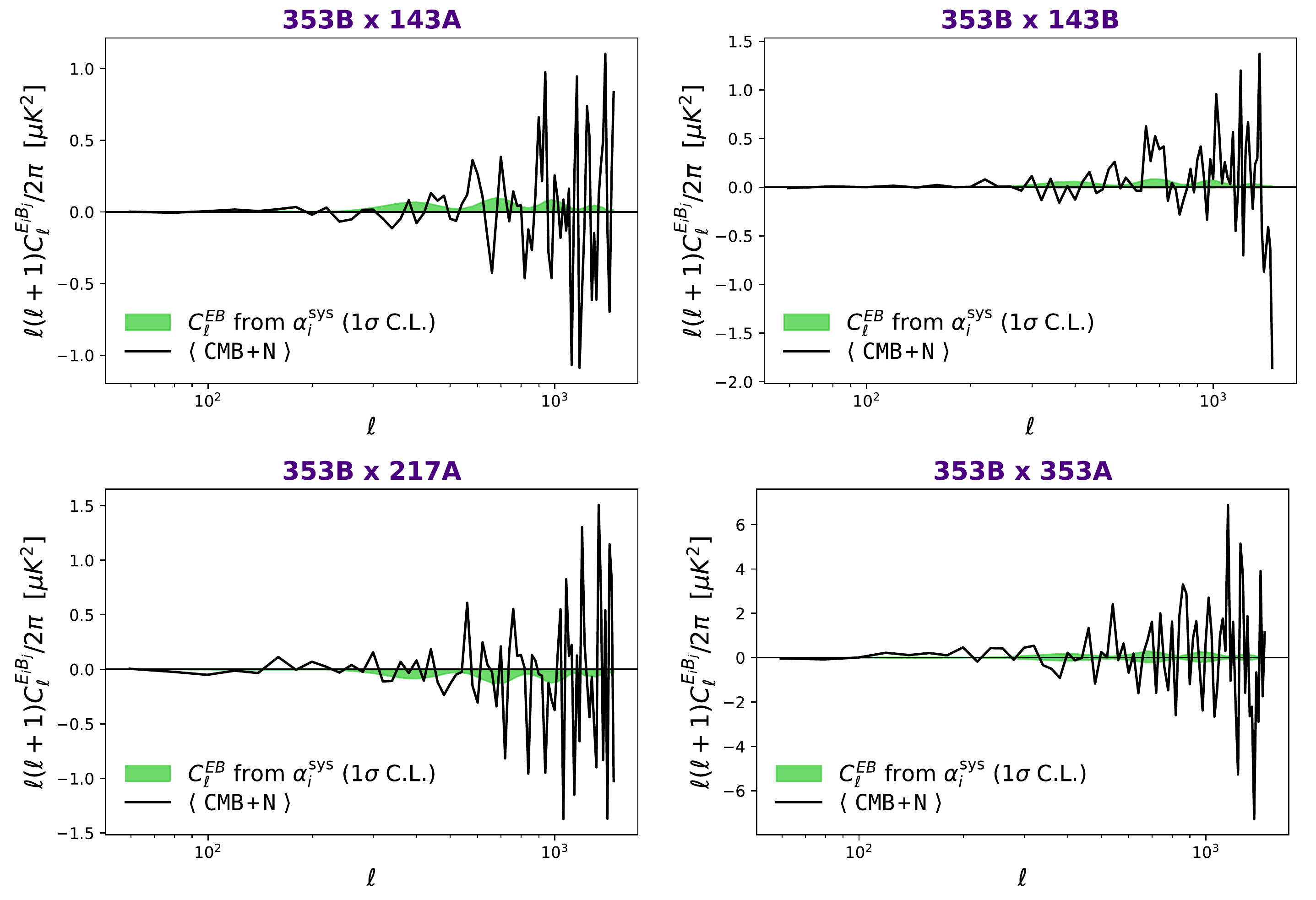}
    \caption{Same as figure~\ref{fig: sys cross pol} but for a selection of detector splits where systematics do not seem to produce a significant spurious $EB$ correlation.}
    \label{fig: sys no sys}
\end{figure}

To understand the origin of the $\alpha_i^\mathrm{sys}$ angles seen in figure~\ref{fig: sys angles bias}, we performed a closer study of the angular power spectra of $\mathtt{CMB+N}$ simulations. In particular, we investigate the origin of the $\alpha_\mathrm{100A}^\mathrm{sys}$ and $\alpha_\mathrm{100B}^\mathrm{sys}$ angles in figure~\ref{fig: sys cross pol}, and that of the $\alpha_\mathrm{143A}^\mathrm{sys}$, $\alpha_\mathrm{143B}^\mathrm{sys}$, and $\alpha_\mathrm{217A}^\mathrm{sys}$ angles in figure~\ref{fig: sys beam leak}. For completeness, in figure~\ref{fig: sys no sys} we also show the angular power spectra of frequency bands where no significant $\alpha_i^\mathrm{sys}$ is found. Black solid lines in figures \ref{fig: sys cross pol}, \ref{fig: sys beam leak}, and \ref{fig: sys no sys} show the mean $C_\ell^{EB}$ angular power spectra for a selection of bands, averaged over the 100 simulations, and binned in uniform bins from $\ell_{\mathrm{min}}=51$ to $\ell_{\mathrm{max}}=1490$ with a spacing of $\Delta\ell=20$. We find no significant difference between mean \texttt{CMB+N} spectra calculated with the \texttt{CO+PS}, \texttt{CO+PS+10\%}, or \texttt{CO+PS+30\%} masks. Thus, we only display the spectra obtained with the \texttt{CO+PS} mask. As demonstrated in figures \ref{fig: sys cross pol} and \ref{fig: sys beam leak}, even in the absence of an $\alpha_i$ miscalibration, \texttt{CMB+N} simulations present a spurious $EB$ correlation between multipoles 200 and 1000 (corresponding roughly to angular scales between 50 and 10 arcmin). These features are more prominent at the lower frequencies, with cross-correlations involving 353GHz showing a mostly uncorrelated $EB$ cross-spectra (see figure~\ref{fig: sys no sys}).

Those spurious $EB$ correlations could be produced by several systematic effects. In general, intensity-to-polarization leakage gives $C_\ell^{EB}\propto C_\ell^{TT}$ at leading order, whereas the cross-polarization effect gives $C_\ell^{EB}\propto C_\ell^{EE}$. A combination of the two would give $C_\ell^{EB}\propto C_\ell^{TE}$. Beam imperfections and mismatches between each detector's optical and electronic responses also lead to a leakage of signal into $EB$. For these simulations that contain only CMB and noise, we calculate the effect that beam leakage has on $EB$:
\begin{equation}
    C_\ell^{EB,\mathrm{BL}}=\omega^2_{\mathrm{pix},\ell}\sum_{XY}W_\ell^{EB,XY}C_\ell^{XY,\Lambda\mathrm{CDM}},
\end{equation}
where $XY\in\{TT,EE,BB,TE\}$, $\omega_{\mathrm{pix},\ell}$ is the pixel window function, and $W_\ell^{EB,XY}$ are the beam-window matrices calculated with \texttt{QuickPol}~\cite{quickpol} specifically for \textit{Planck} beams. To identify which of these effects is most likely to have caused the spurious $EB$ correlations seen in figures \ref{fig: sys cross pol} and \ref{fig: sys beam leak}, we fit the mean \texttt{CMB+N} angular power spectra with the set of models
\begin{align}\label{eq: sys models}\allowdisplaybreaks
   M_b\in\{& A_\mathrm{TT} C_b^{TT,\Lambda\mathrm{CDM}},\\
         & A_\mathrm{EE} C_b^{EE,\Lambda\mathrm{CDM}},\\
         & A_\mathrm{TE} C_b^{TE,\Lambda\mathrm{CDM}},\\
         & A_\mathrm{BL} C_b^{EB,\mathrm{BL}},\\
         & A_\mathrm{TT} C_b^{TT,\Lambda\mathrm{CDM}} + A_\mathrm{EE} C_b^{EE,\Lambda\mathrm{CDM}} + A_\mathrm{TE} C_b^{TE,\Lambda\mathrm{CDM}} + A_\mathrm{BL} C_b^{EB,\mathrm{BL}}\}\label{eq: model combination all sys}
\end{align}
by minimizing a simple $\chi^2$ function: 
\begin{equation}\label{eq: chi cuadrado}
\chi^2_{ij,M} = \sum_b \left( \left\langle C_b^{E_iB_j} \right\rangle - M_b\right)^2 / V_b^{ij}.
\end{equation}
The mean angular power spectra in Eq.~(\ref{eq: chi cuadrado}) are calculated as
\begin{equation}
\left\langle C_b^{XY} \right\rangle =\frac{1}{\Delta\ell}\sum_{\ell \in b} \left\langle C_\ell^{XY}\right\rangle_\mathrm{sim},
\end{equation}
with variance
\begin{equation}
V_b^{ij} = \frac{1}{f_\mathrm{sky}N_\mathrm{sim}\Delta\ell^2} \sum_{\ell\in b}\frac{1}{2\ell+1} \left[ \left\langle C_\ell^{E_iE_i}\right\rangle_\mathrm{sim} \left\langle C_\ell^{B_jB_j} \right\rangle_\mathrm{sim} + \left\langle C_\ell^{E_iB_j} \right\rangle_\mathrm{sim}^2 \right]. \label{eq: variance of the mean}
\end{equation}
Note the $N_\mathrm{sim}^{-1}$ factor in Eq.~(\ref{eq: variance of the mean}), since $\chi^2_{ij,M}$ describes a fit to the mean $EB$ angular power spectra and thus $V_b^{ij}$ is the variance of the mean. 

In figures \ref{fig: sys cross pol} and \ref{fig: sys beam leak}, dashed colored lines show the best fit for each model, with the goodness of fit quantified by the reduced $\chi^2$ included on the bottom-left corner of each plot. To get an intuition of the relative importance of each systematic, we also show on the top-left corner of each plot the values of the $A_\mathrm{TT}$, $A_\mathrm{EE}$, $A_\mathrm{TE}$, and $A_\mathrm{BL}$ amplitudes obtained from the fit to Eq.~(\ref{eq: model combination all sys}). For completeness, the green shaded regions in figures \ref{fig: sys cross pol}, \ref{fig: sys beam leak}, and \ref{fig: sys no sys} show the $1\sigma$ confidence contours of the $EB$ correlation expected from the $\alpha_i^\mathrm{sys}$ angles found in figure~\ref{fig: sys angles bias}. For every combination of $ij$ bands, these contours are generated by plotting the $C_\ell^{E_iB_j} = ( \Sin{4\alpha_j^\mathrm{sys}} C_\ell^{E_iE_j,\mathrm{o}} - \Sin{4\alpha_i^\mathrm{sys}}C_\ell^{B_iB_j,\mathrm{o}}) /(\Cos{4\alpha_i^\mathrm{sys}}+\Cos{4\alpha_j^\mathrm{sys}})$ spectra produced by the angles found in $\mathtt{CMB+N}$ simulations that fall within the 1$\sigma$ confidence ellipse of the correlation between each $\alpha_i^\mathrm{sys}$ $\alpha_j^\mathrm{sys}$ pair. 

The fits in figure~\ref{fig: sys cross pol} suggest the presence of a cross-polarization effect leaking $E$ modes into $B$ modes at 100GHz. This kind of systematic is particularly dangerous since our estimator relies on finding a signal resembling $C_\ell^{EE,\Lambda\mathrm{CDM}}$ in the observed $EB$ correlation to determine both birefringence and miscalibration angles. Moreover, the fit to $A_\mathrm{EE}C_b^{EE,\Lambda\mathrm{CDM}}$ falls perfectly within the $1\sigma$ confidence contours from $\alpha_i^\mathrm{sys}$ angles, confirming that such a cross-polarization effect is indeed the cause of the $\alpha_\mathrm{100A}^\mathrm{sys}$ and $\alpha_\mathrm{100B}^\mathrm{sys}$ angles found in the simulations.

Although the spread in $\chi^2/N_\mathrm{dof}$ from the fits in figure~\ref{fig: sys beam leak} is smaller than that from figure~\ref{fig: sys cross pol}, the fits suggest that beam leakage is the main contribution to the spurious $EB$ correlation seen at those frequencies. Beam leakage has an angular dependence that our estimator cannot reproduce, since it only considers rotations of the observed $EE$ and $BB$ angular power spectra. Nevertheless, the approximate match between the green confidence contours and the mean $EB$ spectra of \texttt{CMB+N} simulations in figure~\ref{fig: sys beam leak} shows how the estimator is trying to accommodate $C_\ell^{EB,\mathrm{BL}}$ as a rotation of $C_\ell^{EE,\Lambda\mathrm{CDM}}$. This limited ability to reproduce the signal from beam leakage leads to the $\alpha_\mathrm{143A}^\mathrm{sys}$, $\alpha_\mathrm{143B}^\mathrm{sys}$, and $\alpha_\mathrm{217A}^\mathrm{sys}$ angles found in the simulations at a lower significance level. 

Finding the presence of these cross-polarization and beam leakage effects is important for understanding all the systematics at play in both simulations and, presumably, the real \textit{Planck} data. Nevertheless, note that the $\alpha_i^\mathrm{sys}$ angles found in \texttt{CMB+N} simulations do not need to agree with the ones found in the data because these simulations do not include the actual (unknown) miscalibration angles present in the data. The $\alpha_i^\mathrm{sys}$ found here would only match the angles found in the data if the orientation of \textit{Planck}'s polarimeters was perfectly calibrated. In this way, the main conclusion to draw from these results is that, even in the presence of such systematics, our methodology is able to correctly capture their effect within the $\alpha_i$ parameters, leaving the measurement of $\beta$ not significantly affected by any of them. The $\langle \bar{\alpha}_\mathrm{sys} \rangle = -0.009^\circ \pm 0.003^\circ$ angle that we find falls well below the corresponding $0.06^\circ$ uncertainty that we have on $\beta$ when simultaneously fitting $\beta$ and $\alpha_i$ (see figure~\ref{fig: sim disp vs fisher Afree}). This observation justifies the decision not to correct the $\beta$ measurement in Ref.~\cite{NPIPE_PRL} for any of the known systematics.

\section{Impact of noise bias}
\label{sec: noise bias}

\begin{figure}[h]
    \begin{minipage}{.49\linewidth}
        \centering 
        \includegraphics[width=\textwidth]{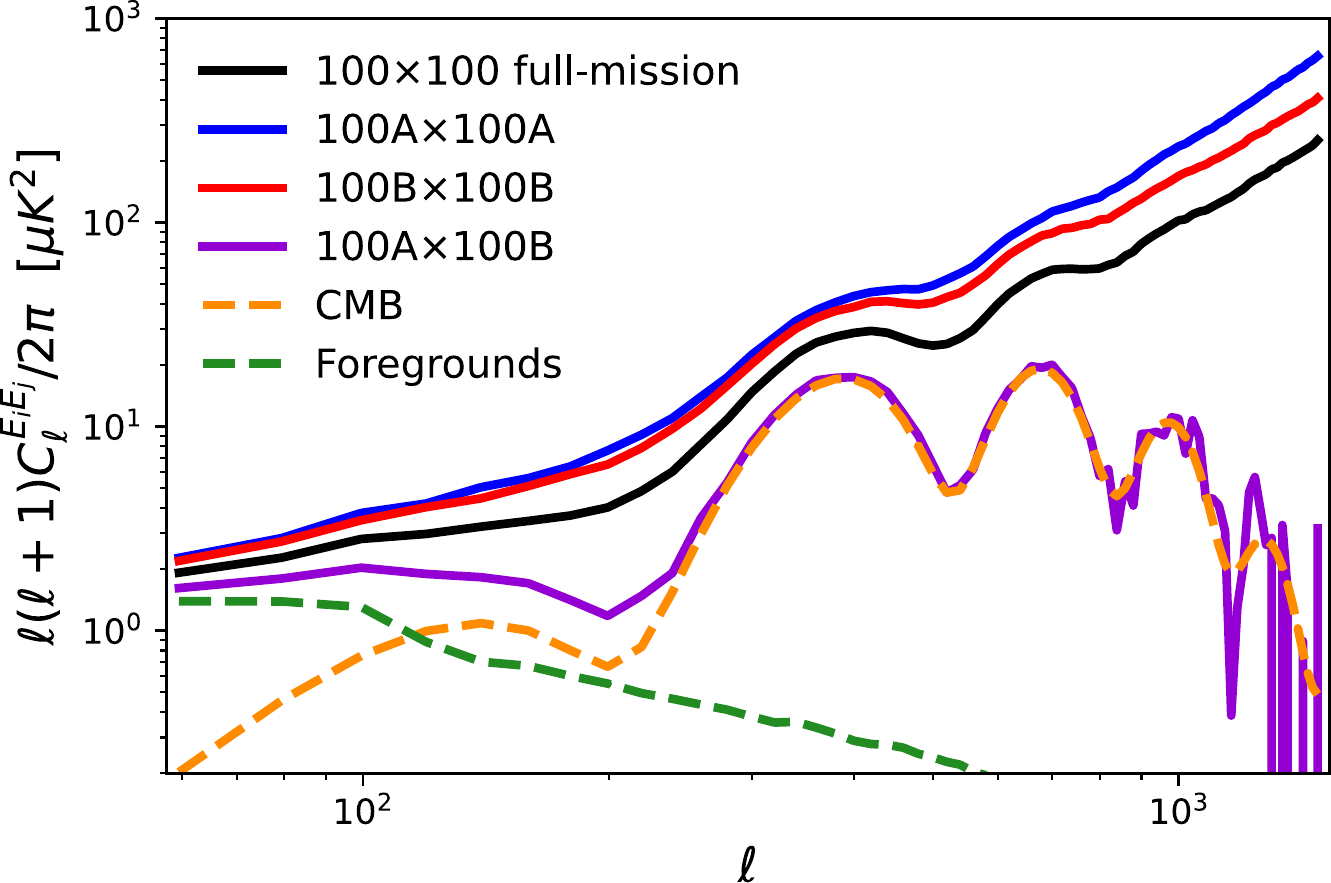}
        \caption{\label{fig: correlated nosie} $EE$ angular power spectra from the auto-correlation of \texttt{NPIPE}'s full-mission simulation at 100GHz (black), and 100A (blue) and 100B (red) detector splits, compared to that from the cross-correlation of A and B detector splits (purple). Foreground (green) and CMB (orange) signals are shown for reference. Spectra were calculated with the \texttt{CO+PS} mask and binned from $\ell\in[51,1490]$ with $\Delta\ell=20$.}
    \end{minipage}
    \hspace{.01\textwidth}
    \begin{minipage}{.49\linewidth}
        \centering 
        \includegraphics[width=\textwidth]{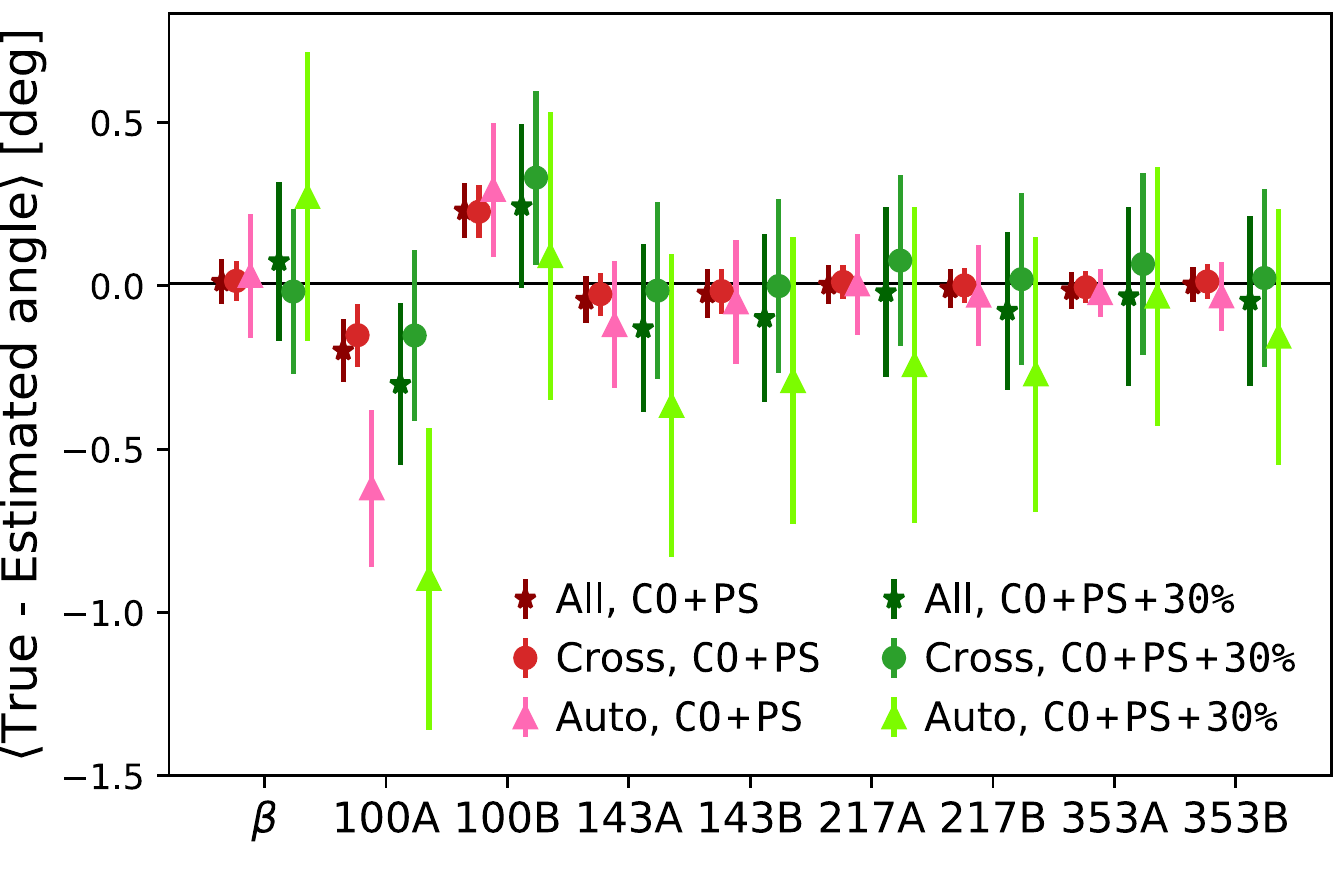}
        \caption{\label{fig: total vs auto vs cross} Bias in the estimation of birefringence and polarization angles from $\mathtt{FG^\alpha+CMB^{\alpha+\beta}+N}$ simulations with estimators that model the foreground $EB$ correlation ($\mathcal{A}$ free), and that use all (stars), cross-spectra-only (circles), and auto-spectra-only (triangles) information. Uncertainties are calculated as the simulations' dispersion. Results are shown for only the smallest and largest Galactic masks considered in this work. }
\end{minipage}
\end{figure}

Although cosmic variance limited for the temperature power spectrum, \textit{Planck}'s polarization noise levels are still relatively high~\cite{Planck_2018_overview}. As seen in figure~\ref{fig: correlated nosie}, the noise bias in the frequency auto-spectra (e.g., 100$\times$100 full-mission) is high enough to obscure most of the CMB and foreground $EE$ signal. Such noise-dominated $EE$ can potentially bias $\beta$ and $\alpha_i$ measurements, since our estimator heavily relies on observed angular power spectra to fit rotation angles and build the covariance matrix. To avoid those biases, frequency auto-spectra were excluded from the analysis of Ref.~\cite{NPIPE_PRL}. Here we quantify the impact of instrumental noise by applying estimators that use information coming from all spectra, only auto-spectra, or only cross-spectra, to $\mathtt{FG^\alpha+CMB^{\alpha+\beta}+N}$ simulations.

Figure~\ref{fig: total vs auto vs cross} shows that measurements derived from auto-spectra-only estimators (light pink and green triangles) lead to higher biases, especially when large Galactic masks are applied. Nevertheless, the higher biases are accompanied by the corresponding increase in the uncertainty, ensuring that the estimates remain compatible with zero within the error bars. When interpreting these results, remember that we expect to recover approximately $-0.2^\circ$ and $0.3^\circ$ values for $\alpha_\mathrm{100A}$ and $\alpha_\mathrm{100B}$, respectively, because of the systematic effects explained in section~\ref{sec:systematics}.

Instrumental noise can be mitigated by cross-correlating different observations of the same signal. \textit{Planck}'s \texttt{NPIPE}~\cite{NPIPE} data release makes this possible by providing A/B detector splits of their frequency maps, which were built from independent subsets of antennas observing at the same frequency. Using cross-spectra, we are able to avoid the noise bias and recover a signal-dominated observed $EE$ spectrum (100A$\times$100B on figure~\ref{fig: correlated nosie}) that improves the estimation of both birefringence and polarization angles. In addition, statistical uncertainties are reduced since the likelihood has $N_{\nu}$ times ($N_\nu-1$ times) more information when all spectra (only cross-spectra) are used~\cite{Minami2020cross}. As seen in figure~\ref{fig: total vs auto vs cross}, the effect of noise is diluted once cross-spectra are included. To further avoid noise bias, we can exclude auto-spectra from the cross-spectra estimator. With respect to the estimator that includes all correlations (dark red and green stars), this cross-spectra-only estimator (red and green circles) reduces the mean value obtained with the \texttt{CO+PS} mask by $30\%$, and that of the \texttt{CO+PS+30\%} mask by $60\%$, while keeping uncertainties compatible within a $5\%$ level.

The superior performance of cross-spectra estimators in noise-dominated experiments like \textit{Planck} was already anticipated in Ref.~\cite{elena_alpha}. Nevertheless, the improvement in polarization noise levels planned for the next generation of CMB experiments will allow for a signal-dominated measurement of $E$ modes without resorting to cross-correlations. Reference \cite{elena_alpha} showed that, for an experiment such as LiteBIRD~\cite{LiteBIRD_PTEP}, auto-spectra-only estimators were more suited for the estimation of miscalibrated polarization angles because of their simpler covariance matrices. We leave the study of the methodology performance in the signal-dominated regime and its application to LiteBIRD for a future work.

\section{Conclusions}
\label{sec:conclusions}

In this work, we have used realistic simulations of \textit{Planck} data to test the impact that Galactic foreground emission and instrumental systematics have on recent birefringence measurements~\cite{PR3_PRL, NPIPE_PRL, johannes, johannes_wmap}. To reduce the computational cost of such an extensive simulation study, we have developed a semi-analytical iterative algorithm that simultaneously calculates birefringence and miscalibrated polarization angles within the small-angle approximation. Our simulation study supports the results presented in Ref.~\cite{NPIPE_PRL}, confirming and highlighting the importance of accounting for dust $EB$ when simultaneously estimating birefringence and miscalibration angles. It also proves that our methodology is robust, not only against the miscalibration of polarization angles, but also against other systematics like intensity-to-polarization leakage, beam leakage, or cross-polarization effects.

We have demonstrated that a model for Galactic foreground emission is needed to calibrate polarization angles and measure cosmic birefringence at the same time. Thus, having a precise characterization of Galactic foregrounds is the most critical aspect of the analysis. For both the simulation study performed here and the application to \textit{Planck}-HFI data presented in Ref.~\cite{NPIPE_PRL}, we adopted the \texttt{Commander} sky model as our foreground model. Although \texttt{Commander} offers one of the best descriptions of thermal dust emission currently available, it still has its limitations. In particular, \texttt{Commander} does not yet provide a signal-dominated template for the foreground $EB$ and it might contain spurious $EB$ correlations through not contemplating the existence of miscalibration angles and the integration of different dust clouds along the line-of-sight in its SED. Our results also lead us to believe that \texttt{Commander} might struggle to reproduce dust emission near the center of the Galactic plane. The comparison of simulations and data tells us that the limited signal-to-noise of the template leads to a $\approx 20\%$ underestimation of the uncertainty of the birefringence angle reported in \cite{NPIPE_PRL}. However, as they are themselves based on the \texttt{Commander} sky model, our simulations do not allow us to quantitatively asses the impact that polarized mixing and miscalibration angles have on the foreground model. Therefore, we leave such a study for a future work.

To overcome these obstacles, a self-consistent end-to-end study encompassing component-separation to birefringence estimation is needed. As demonstrated in Ref.~\cite{elena_alpha} using the \texttt{B-SeCRET} method~\cite{b-secret}, it should be possible to derive a new template of polarized foreground emission free of any spurious $EB$ correlation by adding miscalibration angles to the synchrotron and thermal dust SEDs fitted in Bayesian component-separation analyses. Such a self-consistent study would allow us to correctly propagate uncertainties through the whole pipeline and check the consistency of the birefringence and miscalibration angles obtained at all stages: from frequency maps, to component-separation, and the final clean CMB maps. A better characterization of dust emission beyond the single modified blackbody paradigm~\cite{Vacher_litebird,Vacher_method, Vacher_SED} and high-precision measurements of the CMB polarization from which to derive a signal-dominated template on the full-sky are also required. We believe that such an analysis will allow for an unbiased and reliable measurement of cosmic birefringence in the future.

Finally, here we have limited ourselves to the study of the high-frequency bands of the \textit{Planck} satellite, where thermal dust emission is the main foreground component. Nevertheless, lower frequency bands could be added to the analysis by including an additional template to describe synchrotron radiation. We will explore that extension of the methodology, and its application to the forecasting of LiteBIRD's capabilities and to the analysis of \textit{Planck} and WMAP~\cite{johannes_wmap} data in future works.

\acknowledgments
This research used resources of the National Energy Research Scientific Computing Center (NERSC), a U.S. Department of Energy Office of Science User Facility operated under Contract No.~DE-AC02-05CH11231. Part of the research was carried out at the Jet Propulsion Laboratory, California Institute of Technology, under a contract with the National Aeronautics and Space Administration (80NM0018D0004). PDP acknowledges financial support from the \textit{Formaci\'on del Profesorado Universitario} program of the Spanish Ministerio de Ciencia, Innovaci\'on y Universidades. EdlH acknowledges financial support from the \textit{Concepci\'on Arenal} program of the Universidad de Cantabria. PDP, EMG, PV, BB, and EdlH thank the Spanish Agencia Estatal de Investigaci\'on (AEI, MICIU) for the financial support provided under the projects with references PID2019-110610RB-C21, ESP2017-83921-C2-1-R, and AYA2017-90675-REDC, co-funded with EU FEDER funds, and acknowledge support from Universidad de Cantabria and Consejer{\'i}a de Universidades, Igualdad, Cultura y Deporte del Gobierno de Cantabria via the \textit{Instrumentaci\'on y ciencia de datos para sondear la naturaleza del universo} project, as well as from Unidad de Excelencia Mar\'ia de Maeztu (MDM-2017-0765). JRE acknowledges funding from the European Research Council (ERC) under the Horizon 2020 Research and Innovation Program (Grant agreement No.~819478). The work of YM was supported in part by the Japan Society for the Promotion of Science (JSPS) KAKENHI, Grants No.~JP20K14497. RS and DS acknowledge the support of the Natural Sciences and Engineering Research Council of Canada. The work of EK was supported in part by JSPS KAKENHI Grant No.~JP20H05850 and JP20H05859, and the Deutsche Forschungsgemeinschaft (DFG, German Research Foundation) under Germany's Excellence Strategy - EXC-2094 - 390783311. The Kavli IPMU is supported by World Premier International Research Center Initiative (WPI), MEXT, Japan. We acknowledge the use of \texttt{CAMB}~\cite{camb}, \texttt{HEALPix}~\cite{gorski2005healpix}, \texttt{NaMaster}~\cite{namaster}, \texttt{emcee}~\cite{emcee}, \texttt{corner}~\cite{corner}, \texttt{Matplotlib}~\cite{matplotlib}, and \texttt{Numpy}~\cite{numpy}.

\appendix

\section{Cross-spectra estimator}
\label{sec:appendix cross estimator and covariance}

Starting from Eq.~(\ref{eq: EE EB BE BB correlations}), we build a maximum likelihood estimator that uses the information from all the frequency cross-spectra to simultaneously calculate $\beta$ and $\alpha_i$. In its more general form, the observed $EB$ correlation across different frequency bands is now the rotation of
\begin{align}\label{eq: equation cross-spectra}\allowdisplaybreaks
    C_\ell^{E_iB_j,\mathrm{o}} = & \cfrac{1}{\Cos{4\alpha_i} + \Cos{4\alpha_j}}\Big( \Sin{4\alpha_j} C_\ell^{E_i E_j,\mathrm{o}} - \Sin{4\alpha_i} C_\ell^{B_i B_j,\mathrm{o}} + 2\Cos{2\alpha_i}\Cos{2\alpha_j} {\cal A} C_\ell^{E_i B_j,\mathrm{fg}} \nonumber\\ 
     + & 2 \Sin{2\alpha_i}\Sin{2\alpha_j} {\cal A} C_\ell^{B_i E_j,\mathrm{fg}}  \Big) + \cfrac{ \Sin{4\beta}}{2\Cos{2\alpha_i+2\alpha_j}} \left( C_\ell^{E_iE_j,\Lambda\mathrm{CDM}} - C_\ell^{B_iB_j,\Lambda\mathrm{CDM}}\right).
\end{align}

Analogously to what was done in section~\ref{sec:methodology}, we build a Gaussian likelihood from Eq.~(\ref{eq: equation cross-spectra}) that, within the small-angle approximation, reads
\begin{align}\label{eq: likelihood total small angle}\allowdisplaybreaks
    -2\ln {\cal L} \supset & \sum_{i,j,p,q} \sum_{\ell} \Big[ C_\ell^{E_iB_j,\mathrm{o}} -  2\alpha_j C_\ell^{E_i E_j,\mathrm{o}}
    +2\alpha_i C_\ell^{B_i B_j,\mathrm{o}} - {\cal A} C_\ell^{E_i B_j,\mathrm{fg}} - 2\beta \chi_{ij\ell}^{\Lambda\mathrm{CDM}}\Big]\times \nonumber\\
     & \phantom{\sum } \mathbf{C}_{ijpq\ell}^{-1}\Big[  C_\ell^{E_pB_q,\mathrm{o}} -  2\alpha_q C_\ell^{E_p E_q,\mathrm{o}} +2\alpha_p C_\ell^{B_p B_q,\mathrm{o}} - {\cal A} C_\ell^{E_p B_q,\mathrm{fg}} - 2\beta\chi_{pq\ell}^{\Lambda\mathrm{CDM}}\Big].
\end{align}

In this case, the $\mathbf{C}_{ijpq\ell}$ covariance matrix has $N_\nu^2N_\ell \times N_\nu^2 N_\ell$ elements. Under the approximation in Eq.~(\ref{eq: approx no lxl'}), the $\mathbf{C}_{ijpq\ell}$ covariance matrix in Eq.~(\ref{eq: likelihood total small angle}) can be divided into terms that depend only on the observed, foreground, and CMB spectra, and on their cross-correlations:
\begin{equation}
    \mathbf{C}_{ijpq\ell} = \cfrac{1}{(2\ell +1)f_\mathrm{sky}} \left[ \mathbf{C}_{ijpq\ell}^{\mathrm{o}} + \mathbf{C}_{ijpq\ell}^{\mathrm{CMB}} + \mathbf{C}_{ijpq\ell}^{\mathrm{fg}} + \mathbf{C}_{ijpq\ell}^{\mathrm{CMB*o}} + \mathbf{C}_{ijpq\ell}^{\mathrm{fg*o}} \right].
\end{equation}

Expanding $C_\ell^{X^\mathrm{CMB}Y^\mathrm{o}}$ angular power spectra like $C_\ell^{X^\mathrm{CMB}Y^\mathrm{o}}= \frac{1}{2\ell+1}\sum_{m=-\ell}^{\ell}X_{\ell m}^{\mathrm{CMB}}{Y_{\ell m}^{\mathrm{o}}}^*$, and acknowledging that the spherical harmonic coefficients of the observed signal are a rotation of the CMB and foreground ones as shown in Eq.~(\ref{eq: spherical harmonic coefficients}), the contribution of all CMB-related terms is reduced to
\begin{equation}
   \mathbf{C}_{ijpq\ell}^{\mathrm{CMB}} + \mathbf{C}_{ijpq\ell}^{\mathrm{CMB*o}} = - \cfrac{\SinS{4\beta}b_\ell^i b_\ell^j b_\ell^p b_\ell^q \omega^{4}_{\mathrm{pix},\ell} }{2\Cos{2\alpha_i+2\alpha_j}\Cos{2\alpha_p+2\alpha_q}} \left[ \left(C_\ell^{EE,\Lambda\mathrm{CDM}}\right)^2 + \left(C_\ell^{BB,\Lambda\mathrm{CDM}}\right)^2 \right],
\end{equation}
where $C^{EE,\Lambda\mathrm{CDM}}_\ell$ and $C^{BB,\Lambda\mathrm{CDM}}_\ell$ are the theoretical angular power spectra predicted by $\Lambda$CDM, and the combination of $ij$ frequency bands is specified through the different beam and pixel window functions, $b^i_\ell$ and $\omega_{\mathrm{pix},\ell}$, respectively.

The terms depending on the observed and foreground spectra are calculated as follows:
\begin{align}\label{eq: cov o*o cross}
   \mathbf{C}_{ijpq\ell}^{\mathrm{o}} = &  \phantom{+} C_\ell^{E_i^\mathrm{o}E_p^\mathrm{o}} C_\ell^{B_j^\mathrm{o}B_q^\mathrm{o}} + C_\ell^{E_i^\mathrm{o}B_q^\mathrm{o}} C_\ell^{B_j^\mathrm{o}E_p^\mathrm{o}} \nonumber\\ 
   & + \cfrac{\Sin{4\alpha_p}}{\Cos{4\alpha_p}+\Cos{4\alpha_q}} \left( C_\ell^{E_i^\mathrm{o}B_p^\mathrm{o}} C_\ell^{B_j^\mathrm{o}B_q^\mathrm{o}} + C_\ell^{E_i^\mathrm{o}B_q^\mathrm{o}} C_\ell^{B_j^\mathrm{o}B_p^\mathrm{o}}\right)\nonumber\\
  & - \cfrac{\Sin{4\alpha_q}}{\Cos{4\alpha_p}+\Cos{4\alpha_q}}\left(C_\ell^{E_i^\mathrm{o}E_p^\mathrm{o}} C_\ell^{B_j^\mathrm{o}E_q^\mathrm{o}} + C_\ell^{E_i^\mathrm{o}E_q^\mathrm{o}} C_\ell^{B_j^\mathrm{o}E_p^\mathrm{o}}\right)\nonumber\\ 
  & + \cfrac{\Sin{4\alpha_i}}{\Cos{4\alpha_i}+\Cos{4\alpha_j}}\left( C_\ell^{B_i^\mathrm{o}E_p^\mathrm{o}} C_\ell^{B_j^\mathrm{o}B_q^\mathrm{o}} + C_\ell^{B_i^\mathrm{o}B_q^\mathrm{o}} C_\ell^{B_j^\mathrm{o}E_p^\mathrm{o}} \right)\nonumber\\
  &  - \cfrac{\Sin{4\alpha_j}}{\Cos{4\alpha_i}+\Cos{4\alpha_j}}\left( C_\ell^{E_i^\mathrm{o}E_p^\mathrm{o}} C_\ell^{E_j^\mathrm{o}B_q^\mathrm{o}} + C_\ell^{E_i^\mathrm{o}B_q^\mathrm{o}} C_\ell^{E_j^\mathrm{o}E_p^\mathrm{o}}\right)\nonumber\\
  & +\cfrac{ \Sin{4\alpha_j}\Sin{4\alpha_q}}{[\Cos{4\alpha_i}+\Cos{4\alpha_j}][\Cos{4\alpha_p}+\Cos{4\alpha_q}]} \left( C_\ell^{E_i^\mathrm{o}E_p^\mathrm{o}} C_\ell^{E_j^\mathrm{o}E_q^\mathrm{o}} + C_\ell^{E_i^\mathrm{o}E_q^\mathrm{o}} C_\ell^{E_j^\mathrm{o}E_p^\mathrm{o}} \right)\nonumber\\ 
  & +\cfrac{ \Sin{4\alpha_i}\Sin{4\alpha_p}}{[\Cos{4\alpha_i}+\Cos{4\alpha_j}][\Cos{4\alpha_p}+\Cos{4\alpha_q}]} \left( C_\ell^{B_i^\mathrm{o}B_p^\mathrm{o}} C_\ell^{B_j^\mathrm{o}B_q^\mathrm{o}} + C_\ell^{B_i^\mathrm{o}B_q^\mathrm{o}} C_\ell^{B_j^\mathrm{o}B_p^\mathrm{o}}\right)\nonumber
\end{align}
\begin{align}
  \phantom{\mathbf{C}_{ijpq\ell}^{\mathrm{o}} =}&  -\cfrac{ \Sin{4\alpha_j}\Sin{4\alpha_p}}{[\Cos{4\alpha_i}+\Cos{4\alpha_j}][\Cos{4\alpha_p}+\Cos{4\alpha_q}]} \left( C_\ell^{E_i^\mathrm{o}B_p^\mathrm{o}} C_\ell^{E_j^\mathrm{o}B_q^\mathrm{o}} + C_\ell^{E_i^\mathrm{o}B_q^\mathrm{o}}C_\ell^{E_j^\mathrm{o}B_p^\mathrm{o}}\right)\nonumber\\ 
  &   -\cfrac{ \Sin{4\alpha_i}\Sin{4\alpha_q} }{[\Cos{4\alpha_i}+\Cos{4\alpha_j}][\Cos{4\alpha_p}+\Cos{4\alpha_q}]} \left( C_\ell^{B_i^\mathrm{o}E_p^\mathrm{o}} C_\ell^{B_j^\mathrm{o}E_q^\mathrm{o}} + C_\ell^{B_i^\mathrm{o}E_q^\mathrm{o}} C_\ell^{B_j^\mathrm{o}E_p^\mathrm{o}}\right) 
\end{align}
and
\begin{multline}\allowdisplaybreaks
   \mathbf{C}_{ijpq\ell}^{\mathrm{fg}} = \cfrac{4{\cal A}^2}{[\Cos{4\alpha_i}+\Cos{4\alpha_j}][\Cos{4\alpha_p}+\Cos{4\alpha_q}]}\bigg[\\
   \phantom{+}\Cos{2\alpha_i}\Cos{2\alpha_j}\Cos{2\alpha_p}\Cos{2\alpha_q}\left( C_\ell^{E_i^\mathrm{fg}E_p^\mathrm{fg}} C_\ell^{B_j^\mathrm{fg}B_q^\mathrm{fg}} + C_\ell^{E_i^\mathrm{fg}B_q^\mathrm{fg}} C_\ell^{B_j^\mathrm{fg}E_p^\mathrm{fg}}\right)\\ 
   + \Cos{2\alpha_i}\Cos{2\alpha_j}\Sin{2\alpha_p}\Sin{2\alpha_q}\left(C_\ell^{E_i^\mathrm{fg}B_p^\mathrm{fg}} C_\ell^{B_j^\mathrm{fg}E_q^\mathrm{fg}} + C_\ell^{E_i^\mathrm{fg}E_q^\mathrm{fg}} C_\ell^{B_j^\mathrm{fg}B_p^\mathrm{fg}}\right) \\ 
   + \Sin{2\alpha_i}\Sin{2\alpha_j}\Cos{2\alpha_p}\Cos{2\alpha_q}\left( C_\ell^{B_i^\mathrm{fg}E_p^\mathrm{fg}} C_\ell^{E_j^\mathrm{fg}B_q^\mathrm{fg}} + C_\ell^{B_i^\mathrm{fg}B_q^\mathrm{fg}} C_\ell^{E_j^\mathrm{fg}E_p^\mathrm{fg}} \right)\\ 
   + \Sin{2\alpha_i}\Sin{2\alpha_j}\Sin{2\alpha_p}\Sin{2\alpha_q}\left( C_\ell^{B_i^\mathrm{fg}B_p^\mathrm{fg}} C_\ell^{E_j^\mathrm{fg}E_q^\mathrm{fg}} + C_\ell^{B_i^\mathrm{fg}E_q^\mathrm{fg}} C_\ell^{E_j^\mathrm{fg}B_p^\mathrm{fg}}\right)\bigg]. \hspace{8mm}
\end{multline}

Finally, the cross-correlation between the observed and foreground signals is given as
\begin{align}\allowdisplaybreaks
   \mathbf{C}_{ijpq\ell}^{\mathrm{fg*o}} =& -\cfrac{2{\cal A}\Cos{2\alpha_i}\Cos{2\alpha_j}}{\Cos{4\alpha_i}+\Cos{4\alpha_j}} \left( C_\ell^{E_i^\mathrm{fg}E_p^\mathrm{o}} C_\ell^{B_j^\mathrm{fg}B_q^\mathrm{o}} + C_\ell^{E_i^\mathrm{fg}B_q^\mathrm{o}} C_\ell^{B_j^\mathrm{fg}E_p^\mathrm{o}}\right)\nonumber\\ 
   & -\cfrac{2{\cal A}\Cos{2\alpha_p}\Cos{2\alpha_q}}{\Cos{4\alpha_p}+\Cos{4\alpha_q}} \left( C_\ell^{E_i^\mathrm{o}E_p^\mathrm{fg}} C_\ell^{B_j^\mathrm{o}B_q^\mathrm{fg}} + C_\ell^{E_i^\mathrm{o}B_q^\mathrm{fg}}C_\ell^{B_j^\mathrm{o}E_p^\mathrm{fg}} \right)\nonumber\\   
   & -\cfrac{2{\cal A}\Sin{2\alpha_i}\Sin{2\alpha_j}}{\Cos{4\alpha_i}+\Cos{4\alpha_j}} \left( C_\ell^{B_i^\mathrm{fg}E_p^\mathrm{o}} C_\ell^{E_j^\mathrm{fg}B_q^\mathrm{o}} + C_\ell^{B_i^\mathrm{fg}B_q^\mathrm{o}} C_\ell^{E_j^\mathrm{fg}E_p^\mathrm{o}}\right) \nonumber\\ 
   & -\cfrac{2{\cal A}\Sin{2\alpha_p}\Sin{2\alpha_q}}{\Cos{4\alpha_p}+\Cos{4\alpha_q}} \left( C_\ell^{E_i^\mathrm{o}B_p^\mathrm{fg}} C_\ell^{B_j^\mathrm{o}E_q^\mathrm{fg}} + C_\ell^{E_i^\mathrm{o}E_q^\mathrm{fg}}C_\ell^{B_j^\mathrm{o}B_p^\mathrm{fg}} \right)\nonumber \\ 
   & +\cfrac{2{\cal A}\Cos{2\alpha_p}\Cos{2\alpha_q}\Sin{4\alpha_j}}{[\Cos{4\alpha_i}+\Cos{4\alpha_j}][\Cos{4\alpha_p}+\Cos{4\alpha_q}]} \left( C_\ell^{E_i^\mathrm{o}E_p^\mathrm{fg}} C_\ell^{E_j^\mathrm{o}B_q^\mathrm{fg}} + C_\ell^{E_i^\mathrm{o}B_q^\mathrm{fg}} C_\ell^{E_j^\mathrm{o}E_p^\mathrm{fg}}\right)\nonumber \\
   & +\cfrac{2{\cal A}\Cos{2\alpha_i}\Cos{2\alpha_j}\Sin{4\alpha_q}}{[\Cos{4\alpha_i}+\Cos{4\alpha_j}][\Cos{4\alpha_p}+\Cos{4\alpha_q}]} \left( C_\ell^{E_i^\mathrm{fg}E_p^\mathrm{o}} C_\ell^{B_j^\mathrm{fg}E_q^\mathrm{o}} + C_\ell^{E_i^\mathrm{fg}E_q^\mathrm{o}} C_\ell^{B_j^\mathrm{fg}E_p^\mathrm{o}} \right) \nonumber\\ 
   & +\cfrac{2{\cal A}\Sin{2\alpha_p}\Sin{2\alpha_q}\Sin{4\alpha_j}}{[\Cos{4\alpha_i}+\Cos{4\alpha_j}][\Cos{4\alpha_p}+\Cos{4\alpha_q}]} \left( C_\ell^{E_i^\mathrm{o}B_p^\mathrm{fg}} C_\ell^{E_j^\mathrm{o}E_q^\mathrm{fg}} + C_\ell^{E_i^\mathrm{o}E_q^\mathrm{fg}} C_\ell^{E_j^\mathrm{o}B_p^\mathrm{fg}}\right) \nonumber\\     
   & +\cfrac{2{\cal A}\Sin{2\alpha_i}\Sin{2\alpha_j}\Sin{4\alpha_q}}{[\Cos{4\alpha_i}+\Cos{4\alpha_j}][\Cos{4\alpha_p}+\Cos{4\alpha_q}]} \left( C_\ell^{B_i^\mathrm{fg}E_p^\mathrm{o}} C_\ell^{E_j^\mathrm{fg}E_q^\mathrm{o}} +  C_\ell^{B_i^\mathrm{fg}E_q^\mathrm{o}} C_\ell^{E_j^\mathrm{fg}E_p^\mathrm{o}}\right) \nonumber\\ 
   & -\cfrac{2{\cal A}\Cos{2\alpha_p}\Cos{2\alpha_q}\Sin{4\alpha_i}}{[\Cos{4\alpha_i}+\Cos{4\alpha_j}][\Cos{4\alpha_p}+\Cos{4\alpha_q}]} \left( C_\ell^{B_i^\mathrm{o}E_p^\mathrm{fg}} C_\ell^{B_j^\mathrm{o}B_q^\mathrm{fg}} + C_\ell^{B_i^\mathrm{o}B_q^\mathrm{fg}} C_\ell^{B_j^\mathrm{o}E_p^\mathrm{fg}}\right) \nonumber\\ 
   & -\cfrac{2{\cal A}\Cos{2\alpha_i}\Cos{2\alpha_j}\Sin{4\alpha_p}}{[\Cos{4\alpha_i}+\Cos{4\alpha_j}][\Cos{4\alpha_p}+\Cos{4\alpha_q}]} \left( C_\ell^{E_i^\mathrm{fg}B_p^\mathrm{o}} C_\ell^{B_j^\mathrm{fg}B_q^\mathrm{o}} + C_\ell^{E_i^\mathrm{fg}B_q^\mathrm{o}} C_\ell^{B_j^\mathrm{fg}B_p^\mathrm{o}}\right) \nonumber\\ 
   & -\cfrac{2{\cal A}\Sin{2\alpha_p}\Sin{2\alpha_q}\Sin{4\alpha_i}}{[\Cos{4\alpha_i}+\Cos{4\alpha_j}][\Cos{4\alpha_p}+\Cos{4\alpha_q}]} \left( C_\ell^{B_i^\mathrm{o}B_p^\mathrm{fg}} C_\ell^{B_j^\mathrm{o}E_q^\mathrm{fg}} + C_\ell^{B_i^\mathrm{o}E_q^\mathrm{fg}} C_\ell^{B_j^\mathrm{o}B_p^\mathrm{fg}}\right) \nonumber
\end{align}
\begin{align}\allowdisplaybreaks
   \phantom{\mathbf{C}_{ijpq\ell}^{\mathrm{fg*o}} =} & -\cfrac{2{\cal A}\Sin{2\alpha_i}\Sin{2\alpha_j}\Sin{4\alpha_p}}{[\Cos{4\alpha_i}+\Cos{4\alpha_j}][\Cos{4\alpha_p}+\Cos{4\alpha_q}]} \left( C_\ell^{B_i^\mathrm{fg}B_p^\mathrm{o}} C_\ell^{E_j^\mathrm{fg}B_q^\mathrm{o}} + C_\ell^{B_i^\mathrm{fg}B_q^\mathrm{o}} C_\ell^{E_j^\mathrm{fg}B_p^\mathrm{o}}\right). 
\end{align}

After binning both the angular power spectra and the covariance matrix, the minimization of Eq.~(\ref{eq: likelihood total small angle}) leads to a linear system with the same structure as that of Eq.~(\ref{eq: linear system}), but with $\mathsf{A}_{mn}$ elements that are now
\begin{align} \allowdisplaybreaks
    \Xi =& \phantom{-2} \sum\limits_{i,j,p,q} \sum\limits_{b} C_b^{E_iB_j,\mathrm{fg}} \mathbf{C}_{ijpqb}^{-1} C_b^{E_pB_q,\mathrm{fg}}\label{eq: first term total system}, \\
    Z =& \phantom{-} 2\sum\limits_{i,j,p,q} \sum\limits_{b} C_b^{E_iB_j,\mathrm{fg}} \mathbf{C}_{ijpqb}^{-1} \chi_{pqb}^{\Lambda\mathrm{CDM}}, \label{eq: second term total system}\\
    \Theta =& \phantom{-} 4\sum\limits_{i,j,p,q} \sum\limits_{b} \chi_{ijb}^{\Lambda\mathrm{CDM}} \mathbf{C}_{ijpqb}^{-1} \chi_{pqb}^{\Lambda\mathrm{CDM}}, \label{eq: third term total system}\\
    K_m=& \phantom{-} 2\sum\limits_{i,p,q}\sum\limits_{b} \Big[ C_b^{E_iE_m,\mathrm{o}} \mathbf{C}_{impqb}^{-1} C_b^{E_pB_q,\mathrm{fg}} - C_b^{B_mB_i,\mathrm{o}} \mathbf{C}_{mipqb}^{-1} C_b^{E_pB_q,\mathrm{fg}}\Big], \label{eq: fourth term total system}\\   
    T_m=& \phantom{-} 4\sum\limits_{i,p,q}\sum\limits_{b} \Big[ C_b^{E_iE_m,\mathrm{o}} \mathbf{C}_{impqb}^{-1} \chi_{pqb}^{\Lambda\mathrm{CDM}} - C_b^{B_mB_i,\mathrm{o}} \mathbf{C}_{mipqb}^{-1} \chi_{pqb}^{\Lambda\mathrm{CDM}}\Big], \label{eq: fifth term total system}\\
    \Omega_{mn} =& \phantom{-} 4 \sum\limits_{i,j}\sum\limits_b \Big[  C_b^{E_iE_n,\mathrm{o}} \mathbf{C}_{injmb}^{-1} C_b^{E_jE_m,\mathrm{o}} + C_b^{B_nB_i,\mathrm{o}} \mathbf{C}_{nimjb}^{-1} C_b^{B_mB_j,\mathrm{o}}\Big] \nonumber\\
    & -4\sum\limits_{i,j}\sum\limits_b \Big[ C_b^{B_nB_i,\mathrm{o}} \mathbf{C}_{nijmb}^{-1} C_b^{E_jE_m,\mathrm{o}} + C_b^{B_mB_j,\mathrm{o}} \mathbf{C}_{mjinb}^{-1} C_b^{E_iE_n,\mathrm{o}}\Big]. \label{eq: last term total system}
\end{align}
and $\mathsf{b}_m$ terms
\begin{align}\allowdisplaybreaks
    \xi =& \phantom{2} \sum\limits_{i,j,p,q} \sum\limits_{b} C_b^{E_iB_j,\mathrm{fg}} \mathbf{C}_{ijpqb}^{-1} C_b^{E_pB_q,\mathrm{o}},  \label{eq: first term total independent term}\\
    \theta =& 2\sum\limits_{i,j,p,q} \sum\limits_{b} C_b^{E_iB_j,\mathrm{o}} \mathbf{C}_{ijpqb}^{-1} \chi_{pqb}^{\Lambda\mathrm{CDM}}, \label{eq: second term total independent term}\\
    \omega_m =& 2\sum\limits_{i,p,q}\sum\limits_{b} \Big[ C_b^{E_iE_m,\mathrm{o}} \mathbf{C}_{impqb}^{-1} C_b^{E_pB_q,\mathrm{o}} - C_b^{B_mB_i,\mathrm{o}} \mathbf{C}_{mipqb}^{-1} C_b^{E_pB_q,\mathrm{o}}\Big]\label{eq: last term total independent term},
\end{align}
Once again, the uncertainty in the estimation of the $\mathsf{x}_i=(\mathcal{A}$, $\beta$,$\alpha_i)$ parameters is calculated within the Fisher matrix approximation as $\mathsf{C}^{-1}_{mn}= -\frac{\partial^2{\ln\cal L}}{\partial \mathsf{x}_m \partial \mathsf{x}_n} =\mathsf{A}_{mn}$.

By exploiting the cross-correlation of different frequency bands, the cross-spectra estimator is statistically more powerful than the auto-spectra-only estimator defined in section~\ref{sec:methodology}, due to the sheer increase of available information (from $N_\nu$ to $N_\nu^2$ equations)~\cite{Minami2020cross}. It is also more robust against instrumental noise bias. On the other hand, the greater complexity of the cross-spectra estimator's covariance matrix makes it more prone to suffer from the numerical instabilities that arise from calculating the covariance matrix from observed spectra rather than from theoretical models. As explored in Ref.~\cite{elena_alpha}, such numerical instabilities are mitigated by optimizing the range of multipoles used in the analysis, smoothing the spectra, or binning the covariance matrix.

Note that we can avoid the noise bias contained in frequency auto-spectra with a minimal loss of information if we build an estimator that exclusively uses cross-spectra by explicitly leaving auto-spectra out of the summations in the elements of the linear system. In practice, this is done by changing $\sum_{i,j,p,q}\rightarrow\sum_{i,j\neq i}\sum_{p,q\neq p}$ in Eqs. (\ref{eq: first term total system}), (\ref{eq: second term total system}), (\ref{eq: third term total system}), (\ref{eq: first term total independent term}), and (\ref{eq: second term total independent term}), $\sum_{i,p,q}\rightarrow\sum_{i\neq m}\sum_{p,q\neq p}$ in Eqs. (\ref{eq: fourth term total system}), (\ref{eq: fifth term total system}), and (\ref{eq: last term total independent term}), and $\sum_{i,j}\rightarrow\sum_{i\neq n}\sum_{j\neq m}$ in Eq. (\ref{eq: last term total system}). For this cross-spectra-only estimator, the size of the covariance matrix is reduced to $N_\nu (N_\nu -1)N_\ell \times N_\nu (N_\nu -1) N_\ell$.

\section{Comparison with MCMC sampling}
\label{sec:appendix MCMC compatibility}

\begin{figure}[h]
    \centering
    \includegraphics[width=\linewidth]{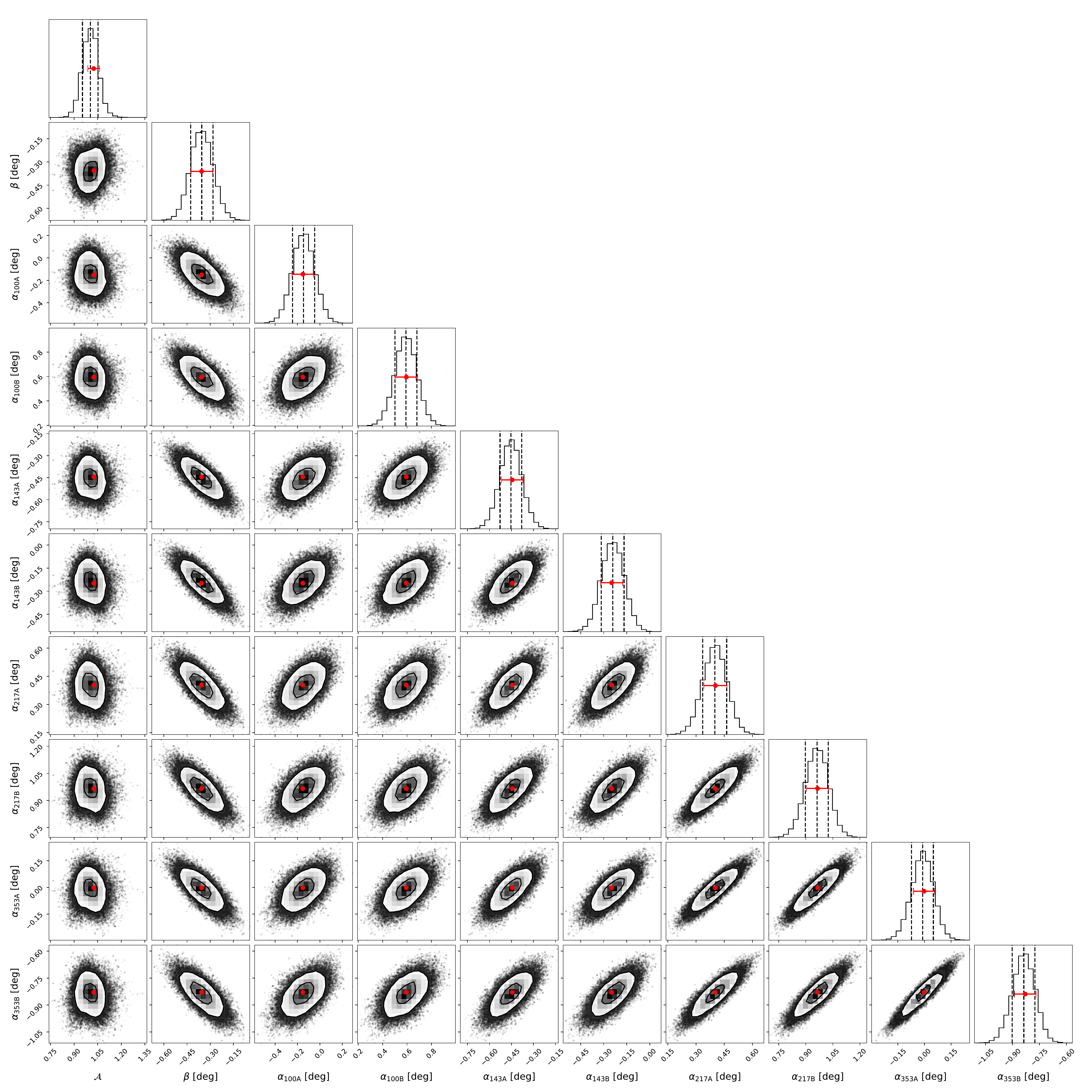}
    \caption{ Posterior distributions obtained from the analysis of one example $\mathtt{FG^\alpha + CMB^{\alpha+\beta} + N}$ simulation using the \texttt{CO+PS} mask. We sample the full likelihood with an MCMC implementation that includes the log-determinant term and corrects for the foreground $EB$ by providing a template. Overlaid in red are the maximum-likelihood solutions and marginalized Fisher uncertainties calculated with our semi-analytical algorithm.}
    \label{fig: corner}
\end{figure}

Here, we briefly compare our semi-analytical algorithm with its counterpart MCMC implementation. By comparing them with the posterior distributions obtained from the MCMC sampling of the full likelihood, figure~\ref{fig: corner} shows that our algorithm is correctly finding the maximum-likelihood solutions and marginalized Fisher uncertainties for all parameters. Those results validate both our iterative approach and the use of the small-angle approximation.

\begin{figure}[h]
    \centering
    \includegraphics[width=0.95\linewidth]{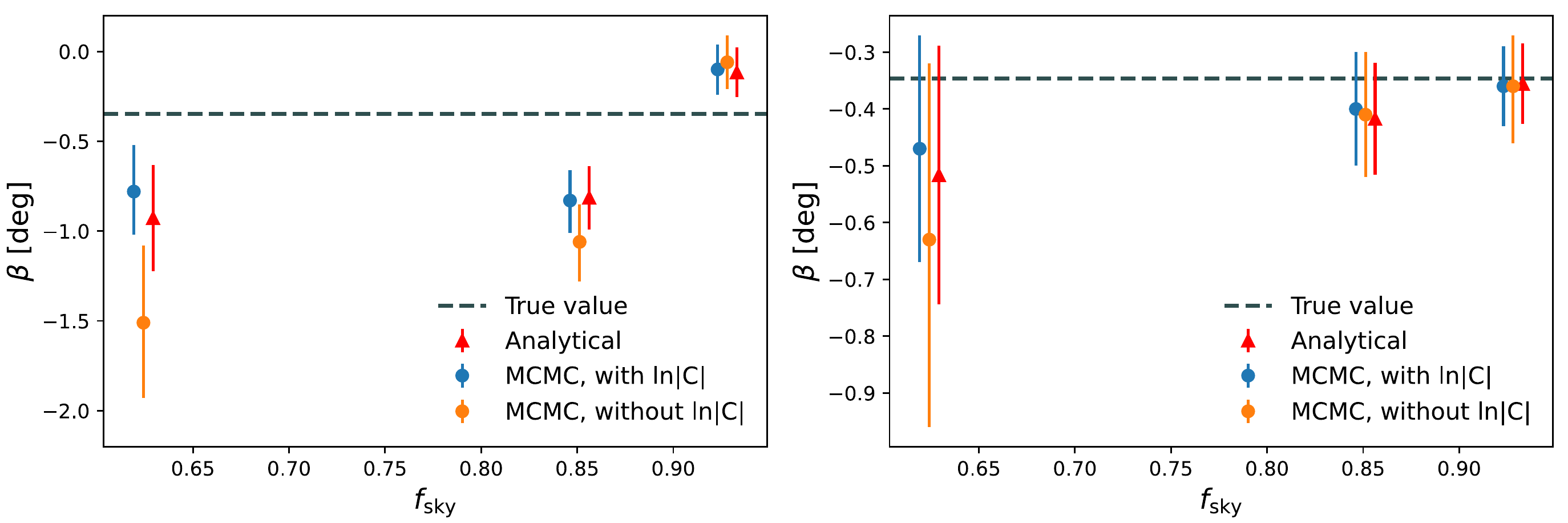}
    \caption{ Maximum-likelihood solutions recovered when sampling the full likelihood with MCMC implementations that include (blue circles) or exclude (orange circles) the log-determinant term, compared to those obtained with our semi-analytical algorithm (red triangles). Results on the left correspond to the analysis of one example $\mathtt{FG^\alpha + CMB^{\alpha+\beta} + N}$ simulation with $\beta=-0.35^\circ$ when the foreground $EB$ is ignored, and those on the right to the analysis of the same simulation when a foreground template is provided.  }
    \label{fig: mcmc vs analytical}
\end{figure}

As discussed in section~\ref{sec:methodology}, the likelihood defined for our estimator should include the $\ln |\mathbf{C}|$ term to ensure that the change in the likelihood's normalization as the free parameters in the covariance matrix vary during the MCMC sampling is taken into account. Therefore, not including the log-determinant can lead to biased posterior distributions. Figure~\ref{fig: mcmc vs analytical} illustrates this effect by showing the maximum-likelihood solutions obtained when including (blue circles) or excluding (orange circles) $\ln |\mathbf{C}|$ from the likelihood, for both the case where foreground $EB$ is ignored (left panel) or accounted for (right panel). The biases produced by ignoring $\ln |\mathbf{C}|$ are more important for smaller sky fractions, and seem to diminish when a template for foreground emission is provided. In both cases, our algorithm yields values (red triangles) that are compatible with those obtained when including $\ln |\mathbf{C}|$, confirming that our iterative approach also accounts for the change in the likelihood's normalization. The uncertainties derived from our algorithm and the MCMC sampling are compatible within a 1\% level for \texttt{CO+PS} and \texttt{CO+PS+10\%} masks, and within 15\% for the \texttt{CO+PS+30\%} mask. The latter discrepancy is of the same order of magnitude as the discrepancy seen between the uncertainties derived from Fisher and the simulations' dispersion in section~\ref{sec:foregrounds}. This suggests that, for large Galactic masks, the Fisher approximation might not be enough to correctly describe posterior distributions.

In terms of computational resources and speed, our semi-analytical algorithm is far superior to the MCMC implementation. Taking the 8 detector splits of \textit{Planck} HFI as a benchmark, and using 60 walkers, running the MCMC sampler parallelized over 64 cores at Cori Haswell\footnote{Cori Haswell nodes have 64 Intel Xeon processors with a 2.3GHz clock rate and a total memory of 128GB per node.} nodes on NERSC takes approximately 4.2s per iteration when the foreground $EB$ contribution is ignored, and 12.9s per iteration when a foreground template is provided. A minimum of around 2500 iterations are needed to obtain fully converged chains, taking from 3 to 9 hours of computation time. In contrast, the semi-analytical algorithm written in plain \texttt{Python} runs on one Cori Haswell core in approximately 7s when the foreground $EB$ contribution is ignored, and 12s when a foreground template is provided. In this sense, our algorithm could be run on any laptop with enough memory to support the volume of data corresponding to the covariance matrix for a given number of frequency bands.

\section{Calculation of the covariance matrix}
\label{sec:appendix covariance matrix}

Here, we offer a detailed calculation of the covariance matrix presented in section~\ref{sec:methodology} for the frequency auto-spectra estimator. We followed the same procedure to calculate the covariance matrix of the cross-spectra estimator presented in appendix~\ref{sec:appendix cross estimator and covariance}. In  Eq.~(\ref{eq: auto cov general expresion}), covariance elements are calculated from the observed angular power spectra as well as the models for both foreground and CMB signals. Therefore, once all the products are expanded, the covariance can be divided into terms that depend only on the observed, foreground, and CMB spectra, and on their cross-correlations:
\begin{equation}\label{eq: total covariance autos}
    \mathbf{C}_{ij\ell} = \cfrac{1}{(2\ell +1)f_{\mathrm{sky}}} \left[ \mathbf{C}_{ij\ell}^{\mathrm{o}} + \mathbf{C}_{ij\ell}^{\mathrm{CMB}} + \mathbf{C}_{ij\ell}^{\mathrm{fg}} + \mathbf{C}_{ij\ell}^{\mathrm{CMB*o}} + \mathbf{C}_{ij\ell}^{\mathrm{fg*o}}  \right],
\end{equation}
where we have already assumed that correlations between the foreground and CMB signals are negligible\footnote{Although chance correlations between foreground and CMB signals can be important on a realization-by-realization basis, they are subdominant at the angular scales of interest for this work ($\ell>50$).} ($\mathbf{C}_{ij\ell}^{\mathrm{CMB*fg}}=0$), and included the sky fraction factor $f_{\mathrm{sky}}$ to account for partial sky coverage. 

Applying Eq.~(\ref{eq: approx no lxl'}), $\mathbf{C}_{ij\ell}^{\mathrm{o}}$ and $\mathbf{C}_{ij\ell}^{\mathrm{fg}}$ terms are calculated as follows:
\begin{align}\label{eq: cov o*o auto}\allowdisplaybreaks
   \mathbf{C}_{ij\ell}^{\mathrm{o}} = & \phantom{+} C_\ell^{E_i^\mathrm{o}E_j^\mathrm{o}} C_\ell^{B_i^\mathrm{o}B_j^\mathrm{o}} + C_\ell^{E_i^\mathrm{o}B_j^\mathrm{o}} C_\ell^{B_i^\mathrm{o}E_j^\mathrm{o}} \nonumber\\ 
    & + \Tan{4\alpha_j} \left[ C_\ell^{E_i^\mathrm{o}B_j^\mathrm{o}} C_\ell^{B_i^\mathrm{o}B_j^\mathrm{o}} - C_\ell^{E_i^\mathrm{o}E_j^\mathrm{o}} C_\ell^{B_i^\mathrm{o}E_j^\mathrm{o}}\right] \nonumber\\
    & + \Tan{4\alpha_i} \left[ C_\ell^{B_i^\mathrm{o}E_j^\mathrm{o}} C_\ell^{B_i^\mathrm{o}B_j^\mathrm{o}} - C_\ell^{E_i^\mathrm{o}E_j^\mathrm{o}} C_\ell^{E_i^\mathrm{o}B_j^\mathrm{o}}\right] \nonumber\\ 
    & + \cfrac{\Tan{4\alpha_i}\Tan{4\alpha_j}}{2} \left[ \left(C_\ell^{E_i^\mathrm{o}E_j^\mathrm{o}}\right)^2 + \left( C_\ell^{B_i^\mathrm{o}B_j^\mathrm{o}}\right)^2 - \left(C_\ell^{E_i^\mathrm{o}B_j^\mathrm{o}}\right)^2 - \left(C_\ell^{B_i^\mathrm{o}E_j^\mathrm{o}}\right)^2 \right]  
\end{align}
and
\begin{equation}\allowdisplaybreaks
   \mathbf{C}_{ij\ell}^{\mathrm{fg}} = \cfrac{{\cal A}^2}{\Cos{4\alpha_i}\Cos{4\alpha_j}}\left[ C_\ell^{E_i^\mathrm{fg}E_j^\mathrm{fg}} C_\ell^{B_i^\mathrm{fg}B_j^\mathrm{fg}} + C_\ell^{E_i^\mathrm{fg}B_j^\mathrm{fg}} C_\ell^{B_i^\mathrm{fg}E_j^\mathrm{fg}} \right].
\end{equation}

The cross-correlation between the observed signal and the foreground model is given as
\begin{align}\label{eq: cov fg*o auto}\allowdisplaybreaks
   \mathbf{C}_{ij\ell}^{\mathrm{fg*o}} = & - \cfrac{{\cal A}}{\Cos{4\alpha_i}} \left[C_\ell^{E_i^\mathrm{fg}E_j^\mathrm{o}} C_\ell^{B_i^\mathrm{fg}B_j^\mathrm{o}} + C_\ell^{B_i^\mathrm{fg}E_j^\mathrm{o}} C_\ell^{E_i^\mathrm{fg}B_j^\mathrm{o}} \right] \nonumber \\
   & - \cfrac{{\cal A}}{\Cos{4\alpha_j}} \left[ C_\ell^{E_i^\mathrm{o}E_j^\mathrm{fg}} C_\ell^{B_i^\mathrm{o}B_j^\mathrm{fg}} + C_\ell^{E_i^\mathrm{o}B_j^\mathrm{fg}} C_\ell^{B_i^\mathrm{o}E_j^\mathrm{fg}} \right] \nonumber\\
   & + \cfrac{{\cal A}\Tan{4\alpha_j}}{\Cos{4\alpha_i}} \left[ C_\ell^{E_i^\mathrm{fg}E_j^\mathrm{o}} C_\ell^{B_i^\mathrm{fg}E_j^\mathrm{o}} -C_\ell^{E_i^\mathrm{fg}B_j^\mathrm{o}} C_\ell^{B_i^\mathrm{fg}B_j^\mathrm{o}}\right]\nonumber\\
   & +\cfrac{{\cal A}\Tan{4\alpha_i}}{\Cos{4\alpha_j}} \left[ C_\ell^{E_i^\mathrm{o}E_j^\mathrm{fg}} C_\ell^{E_i^\mathrm{o}B_j^\mathrm{fg}} -C_\ell^{B_i^\mathrm{o}E_j^\mathrm{fg}} C_\ell^{B_i^\mathrm{o}B_j^\mathrm{fg}}\right]. 
\end{align}
If we had a theoretical model for the foreground angular power spectra, or accepted the \texttt{Commander} sky model as an exact representation of the polarized foreground emission on the sky, we could further expand the $C_\ell^{X^\mathrm{fg}Y^\mathrm{o}}$ terms in Eq.~(\ref{eq: cov fg*o auto}) by acknowledging that the spherical harmonic coefficients of the observed signal are a rotation of the CMB and foreground ones (see Eq.~(\ref{eq: spherical harmonic coefficients})). Therefore, when calculating $C_\ell^{X^\mathrm{fg}Y^\mathrm{o}}=(2\ell+1)^{-1} \sum_{m=-\ell}^{\ell} {X_{\ell m}^{\mathrm{fg}}}^{\phantom{*}}{Y_{\ell m}^{\mathrm{o}}}^*$ for any given pair of frequency bands, we will obtain a rotation of $C_\ell^{E^\mathrm{fg}E^\mathrm{fg}}$, $C_\ell^{E^\mathrm{fg}B^\mathrm{fg}}$, and $C_\ell^{B^\mathrm{fg}B^\mathrm{fg}}$. Instead, as we are treating \texttt{Commander} as an approximate model, in this work we calculate the $C_\ell^{X^\mathrm{fg}Y^\mathrm{o}}$ correlations between the observed maps and \texttt{Commander} templates to account for any possible mismodeling of the foreground emission. 

On the other hand, for the CMB we do expand the corresponding $C_\ell^{X^\mathrm{CMB}Y^\mathrm{o}}$ terms. In that case, the contribution from CMB-related terms to the covariance is
\begin{align}
   \mathbf{C}_{ij\ell}^{\mathrm{CMB}} =&  \phantom{-} \cfrac{\SinS{4\beta} }{2\Cos{4\alpha_i}\Cos{4\alpha_j}} \left(b_\ell^i b_\ell^j\right)^2 \omega^{4}_{\mathrm{pix},\ell} \left[ \left(C_\ell^{EE,\Lambda\mathrm{CDM}}\right)^2 + \left(C_\ell^{BB,\Lambda\mathrm{CDM}}\right)^2 \right], \label{eq: cov cmb*cmb auto}\\
   \mathbf{C}_{ij\ell}^{\mathrm{CMB*o}} =& - \cfrac{\SinS{4\beta} }{\Cos{4\alpha_i}\Cos{4\alpha_j}} \left(b_\ell^i b_\ell^j\right)^2 \omega^{4}_{\mathrm{pix},\ell} \left[ \left(C_\ell^{EE,\Lambda\mathrm{CDM}}\right)^2 + \left(C_\ell^{BB,\Lambda\mathrm{CDM}}\right)^2 \right], \label{eq: cov cmb*o auto}
\end{align}
where $C_\ell^{EE,\Lambda\mathrm{CDM}}$ and $C_\ell^{BB,\Lambda\mathrm{CDM}}$ are the theoretical angular power spectra predicted by $\Lambda$CDM, and the combination of $ij$ frequency bands is specified through the different beam and pixel window functions, $b_\ell^i$ and $\omega_{\mathrm{pix},\ell}$, respectively.

Note that the covariance matrix is a block matrix composed of $N_\nu\times N_\nu$ diagonal $N_\ell\times N_\ell$ boxes since we are not considering $\ell$-to-$\ell'$ correlations. Hence, our algorithm can be optimized by reordering the $\mathbf{C}_{ij\ell\ell'}$ terms of the covariance into $N_\ell \times N_\ell$ boxes of $N_\nu \times N_\nu$ elements to form a block diagonal matrix whose inverse is calculated by independently inverting each of its blocks. This leads to a faster implementation, since inverting $N_\ell$ $N_\nu \times N_\nu$ matrices is faster than inverting one big $N_\nu N_\ell\times N_\nu N_\ell$ matrix, especially when dealing with a large number of frequency bands.

\section{Modeling Galactic foregrounds in the covariance matrix}
\label{sec:appendix foreground EB in cov}

In section~\ref{sec:foregrounds}, we saw that the statistical uncertainties in the estimation of $\beta$ and $\alpha_i$ decreased when the foreground $EB$ was included in the model ($\mathcal{A}\neq 0$). Although counter-intuitive at first, here we will explain why the inclusion of a template for Galactic foreground emission in the covariance matrix leads to a reduction of the total covariance that produces those smaller uncertainties. For simplicity, we will perform these calculations for the frequency auto-spectra estimator.

To determine the role of foregrounds in the covariance matrix, we can assume that our foreground template is a faithful representation of the foreground emission in the sky and expand the $C_\ell^{X^\mathrm{fg}Y^\mathrm{o}}$ terms in Eq.~(\ref{eq: cov fg*o auto}) as a rotation of $C_\ell^{E^\mathrm{fg}E^\mathrm{fg}}$, $C_\ell^{E^\mathrm{fg}B^\mathrm{fg}}$, and $C_\ell^{B^\mathrm{fg}B^\mathrm{fg}}$. Under these conditions, the contribution of the foreground template to the covariance is
\begin{align} \allowdisplaybreaks
   \mathbf{C}_{ij\ell}^{\mathrm{fg}} =& \phantom{-} \cfrac{ \mathcal{A}^2 }{ \Cos{4\alpha_i}\Cos{4\alpha_j} } \left[ C_\ell^{E_iE_j,\mathrm{fg}} C_\ell^{B_iB_j,\mathrm{fg}} +C_\ell^{E_iB_j,\mathrm{fg}} C_\ell^{B_iE_j,\mathrm{fg}} \right],\\
   \mathbf{C}_{ij\ell}^{\mathrm{fg*o}} =& - \cfrac{ 2\mathcal{A} }{ \Cos{4\alpha_i}\Cos{4\alpha_j} } \left[ C_\ell^{E_iE_j,\mathrm{fg}} C_\ell^{B_iB_j,\mathrm{fg}} +C_\ell^{E_iB_j,\mathrm{fg}} C_\ell^{B_iE_j,\mathrm{fg}} \right].
\end{align}
From these terms alone we can already see that if the template offers a good enough representation of the foreground emission in the sky, then $\mathcal{A}\approx1$, and $\mathbf{C}_{ij\ell}^{\mathrm{fg}} + \mathbf{C}_{ij\ell}^{\mathrm{fg*o}}$ becomes a negative contribution to the total covariance.

We can also expand the $C_\ell^{X^\mathrm{o}Y^\mathrm{o}}$ angular power spectra in Eq.~(\ref{eq: cov o*o auto}) by explicitly calculating the correlations between the rotated foreground and CMB components, as written in Eq.~(\ref{eq: spherical harmonic coefficients}). With that we obtain
\begin{align*}\allowdisplaybreaks
   \mathbf{C}^{\mathrm{o}}_{ij\ell} = & \phantom{+} \cfrac{1}{\Cos{4\alpha_i}\Cos{4\alpha_j}} \left[ C_\ell^{E_iE_j,\mathrm{fg}} C_\ell^{B_iB_j,\mathrm{fg}} + C_\ell^{E_iB_j,\mathrm{fg}} C_\ell^{B_iE_j,\mathrm{fg}}\right]\\
   & + \cfrac{\SinS{4\beta}}{2\Cos{4\alpha_i}\Cos{4\alpha_j}}\left[ \left( C_\ell^{E_iE_j,\Lambda\mathrm{CDM}}\right)^2 +\left( C_\ell^{B_iB_j,\Lambda\mathrm{CDM}}\right)^2 \right]\\
   & + \cfrac{\Sin{4\beta}\CosS{2\alpha_i+2\alpha_j}}{\Cos{4\alpha_i}\Cos{4\alpha_j}} \left[ C_\ell^{E_iB_j,\mathrm{fg}} + C_\ell^{B_iE_j,\mathrm{fg}}\right]\left[C_\ell^{E_iE_j,\Lambda\mathrm{CDM}} - C_\ell^{B_iB_j,\Lambda\mathrm{CDM}}\right]\\
   & + \cfrac{\SinS{2\beta}}{\Cos{4\alpha_i}\Cos{4\alpha_j}} \left[ C_\ell^{E_iE_j,\mathrm{fg}} C_\ell^{E_iE_j,\Lambda\mathrm{CDM}} + C_\ell^{B_iB_j,\mathrm{fg}} C_\ell^{B_iB_j,\Lambda\mathrm{CDM}}\right]\\
   & + \cfrac{\CosS{2\beta}}{\Cos{4\alpha_i}\Cos{4\alpha_j}} \left[ C_\ell^{E_iE_j,\mathrm{fg}} C_\ell^{B_iB_j,\Lambda\mathrm{CDM}} + C_\ell^{B_iB_j,\mathrm{fg}} C_\ell^{E_iE_j,\Lambda\mathrm{CDM}}\right] 
\end{align*}
\begin{equation}\label{eq: cov o*o auto expandido}\allowdisplaybreaks
   + \cfrac{\SinS{4\beta}}{\Cos{4\alpha_i}\Cos{4\alpha_j}}C_\ell^{E_iE_j,\Lambda\mathrm{CDM}} C_\ell^{B_iB_j,\Lambda\mathrm{CDM}}. \hspace{4cm}
\end{equation}
Adding all contributions, the total covariance in Eq.~(\ref{eq: total covariance autos}) is
\begin{align}\label{eq: expanded total cov auto}\allowdisplaybreaks
   (2\ell+1)f_{\mathrm{sky}}\mathbf{C}_{ij\ell} = & \phantom{+} \cfrac{1+\mathcal{A}(\mathcal{A}-2)}{\Cos{4\alpha_i}\Cos{4\alpha_j}} \left[ C_\ell^{E_iE_j,\mathrm{fg}} C_\ell^{B_iB_j,\mathrm{fg}} + C_\ell^{E_iB_j,\mathrm{fg}} C_\ell^{B_iE_j,\mathrm{fg}}\right] \nonumber\\
   & + \cfrac{\Sin{4\beta}\CosS{2\alpha_i+2\alpha_j}}{\Cos{4\alpha_i}\Cos{4\alpha_j}} \left[ C_\ell^{E_iB_j,\mathrm{fg}} + C_\ell^{B_iE_j,\mathrm{fg}}\right]\left[C_\ell^{E_iE_j,\Lambda\mathrm{CDM}} - C_\ell^{B_iB_j,\Lambda\mathrm{CDM}}\right] \nonumber\\
   & + \cfrac{\SinS{2\beta}}{\Cos{4\alpha_i}\Cos{4\alpha_j}} \left[ C_\ell^{E_iE_j,\mathrm{fg}} C_\ell^{E_iE_j,\Lambda\mathrm{CDM}} + C_\ell^{B_iB_j,\mathrm{fg}} C_\ell^{B_iB_j,\Lambda\mathrm{CDM}}\right] \nonumber\\
   & + \cfrac{\CosS{2\beta}}{\Cos{4\alpha_i}\Cos{4\alpha_j}} \left[ C_\ell^{E_iE_j,\mathrm{fg}} C_\ell^{B_iB_j,\Lambda\mathrm{CDM}} + C_\ell^{B_iB_j,\mathrm{fg}} C_\ell^{E_iE_j,\Lambda\mathrm{CDM}}\right] \nonumber\\
   & + \cfrac{\SinS{4\beta}}{\Cos{4\alpha_i}\Cos{4\alpha_j}}C_\ell^{E_iE_j,\Lambda\mathrm{CDM}} C_\ell^{B_iB_j,\mathrm{cmb}},   
\end{align}
where the second term from Eq.~(\ref{eq: cov o*o auto expandido}) gets cancelled by the sum of $\mathbf{C}_{ij\ell}^{\mathrm{CMB}}$ and $\mathbf{C}_{ij\ell}^{\mathrm{CMB*o}}$ from Eqs.~(\ref{eq: cov cmb*cmb auto}) and (\ref{eq: cov cmb*o auto}) as long as our theoretical model for the CMB angular power spectra is accurate enough. At high frequencies, Galactic foreground emission dominates over that of the CMB (especially at the largest scales), making the first term in Eq.~(\ref{eq: expanded total cov auto}) one of the main contributions to the total covariance. That term remains present if we do not account for the foreground $EB$ correlation ($\mathcal{A}=0$) in our likelihood, but gets cancelled when we use a good enough foreground template ($\mathcal{A}\approx 1$) to correct for it. This cancellation explains why including the foreground template leads to a reduction of the statistical uncertainties associated with our measurements.

It is also worth noting that $\mathbf{C}_{ij\ell}^{\mathrm{fg*o}}$ is the term responsible for the reduction in the covariance matrix. Thus, the inclusion of foregrounds will indeed lead to an increase in statistical uncertainties if we ignored the correlations between template and data. The same happens with $\mathbf{C}_{ij\ell}^{\mathrm{CMB}}$ and  $\mathbf{C}_{ij\ell}^{\mathrm{CMB*o}}$ correlations.

\bibliographystyle{JHEP}
\bibliography{references}

\end{document}